\documentclass[twocolumn]{aastex62}
\usepackage{longtable}
\usepackage{natbib}
\usepackage{latexsym,amsmath,amssymb,amsfonts,hyperref,chngcntr} 
\newcommand{\hi}{\mbox{H\,{\sc i}}}

\newcommand{\kms}{km\,s$^{-1}$}
\newcommand{\cmsq}{cm$^{-2}$}

\newcommand{\mgii}{\mbox{Mg\,{\sc ii}}} 
\newcommand{\mgiia}{\mbox{Mg\,{\sc ii}$\lambda$2796}}

\newcommand{\mgiiab}{\mbox{Mg\,{\sc ii}$\lambda\lambda$2796,2803}}

\submitjournal{ApJ}
\shorttitle{Blind \hi\ and OH survey: MALS and uGMRT} 
\shortauthors{Gupta et al.}
\begin{document}

\title{Blind \hi\ and OH absorption line search: first results with MALS and uGMRT processed using ARTIP} 

\correspondingauthor{N. Gupta}
\email{ngupta@iucaa.in}

\author{N. Gupta},  
\affil{Inter-University Centre for Astronomy and Astrophysics, Post Bag 4, Ganeshkhind, Pune 411 007, India}

\author{P. Jagannathan}  
\affil{National Radio Astronomy Observatory, Socorro, NM 87801, USA}

\author{R. Srianand},  
\affil{Inter-University Centre for Astronomy and Astrophysics, Post Bag 4, Ganeshkhind, Pune 411 007, India}

\author{S. Bhatnagar}  
\affil{National Radio Astronomy Observatory, Socorro, NM 87801, USA}

\author{P. Noterdaeme}  
\affil{Institut d'astrophysique de Paris, UMR 7095, CNRS-SU, 98bis bd Arago, 75014  Paris, France}

\author{F. Combes}  
\affil{ Observatoire de Paris, Coll\`ege de France, PSL University, Sorbonne University, CNRS, LERMA, Paris, France}

\author{P. Petitjean}  
\affil{Institut d'astrophysique de Paris, UMR 7095, CNRS-SU, 98bis bd Arago, 75014  Paris, France}

\author{J. Jose}  
\affil{ThoughtWorks Technologies India Private Limited, Yerawada, Pune 411 006, India}

\author{S. Pandey}  
\affil{ThoughtWorks Technologies India Private Limited, Yerawada, Pune 411 006, India}

\author{C. Kaski}  
\affil{ThoughtWorks Technologies India Private Limited, Yerawada, Pune 411 006, India}

\author{A. J. Baker}  
\affil{Department of Physics and Astronomy, Rutgers, the State University of New Jersey, 136 Frelinghuysen Road, Piscataway, NJ 08854-8019, USA}

\author{S. A. Balashev}  
\affil{Ioffe Institute, 26 Politeknicheskaya st., St. Petersburg, 194021, Russia}

\author{E. Boettcher}  
\affil{Department of Astronomy \& Astrophysics, The University of Chicago, 5640 S Ellis Ave., Chicago, IL 60637, USA}

\author{H.-W. Chen}  
\affil{Department of Astronomy \& Astrophysics, The University of Chicago, 5640 S Ellis Ave., Chicago, IL 60637, USA}

\author{C. Cress}  
\affil{Department of Physics and Astronomy, University of the Western Cape, Robert Sobukwe Road, Bellville, 7535, South Africa}

\author{R. Dutta}  
\affil{Institute for Computational Cosmology, Durham University, South Road, Durham, DH1 3LE, UK}
\affil{Centre for Extragalactic Astronomy, Durham University, South Road, Durham, DH1 3LE, UK}

\author{S. Goedhart}  
\affil{South African Radio Astronomy Observatory, 2 Fir Street, Black River Park, Observatory 7925, South Africa}

\author{G. Heald}  
\affil{CSIRO Astronomy and Space Science, PO Box 1130, Bentley, WA 6012, Australia}

\author{G. I. G. J\'ozsa}  
\affil{South African Radio Astronomy Observatory, 2 Fir Street, Black River Park, Observatory 7925, South Africa}
\affil{Department of Physics and Electronics, Rhodes University, P.O. Box 94, Makhanda, 6140, South Africa}
\affil{Argelander-Institut f\"ur Astronomie, Auf dem H\"ugel 71, D-53121 Bonn, Germany}

\author{E. Kamau}  
\affil{South African Radio Astronomy Observatory, 2 Fir Street, Black River Park, Observatory 7925, South Africa}
\affil{Department of Physics and Electronics, Rhodes University, P.O. Box 94, Makhanda, 6140, South Africa}

\author{P. Kamphuis}  
\affil{Ruhr University Bochum, Faculty of Physics and Astronomy, Astronomical Institute, 44780 Bochum, Germany}

\author{J. Kerp}  
\affil{Argelander-Institut f\"ur Astronomie, Auf dem H\"ugel 71, D-53121 Bonn, Germany}

\author{H.-R. Kl\"ockner}  
\affil{Max-Planck-Institut f\"ur Radioastronomie, Auf dem H\"ugel 69, D-53121 Bonn, Germany}

\author{K. Knowles}  
\affil{Astrophysics Research Centre and School of Mathematics, Statistics and Computer Science, University of KwaZulu-Natal, Durban 4041, South Africa}

\author{V. Krishnan}  
\affil{South African Radio Astronomy Observatory, 2 Fir Street, Black River Park, Observatory 7925, South Africa}

\author{J-.K. Krogager}  
\affil{Institut d'astrophysique de Paris, UMR 7095, CNRS-SU, 98bis bd Arago, 75014  Paris, France}

\author{V. P. Kulkarni}  
\affil{Department of Physics and Astronomy, University of South Carolina, Columbia, SC 29208, USA}

\author{E. Momjian}  
\affil{National Radio Astronomy Observatory, Socorro, NM 87801, USA}

\author{K. Moodley}  
\affil{Astrophysics Research Centre and School of Mathematics, Statistics and Computer Science, University of KwaZulu-Natal, Durban 4041, South Africa }

\author{S. Passmoor}  
\affil{South African Radio Astronomy Observatory, 2 Fir Street, Black River Park, Observatory 7925, South Africa}

\author{A. Schr\"oeder}  
\affil{South African Astronomical Observatory, P.O. Box 9, Observatory 7935, Cape Town, South Africa}

\author{S. Sekhar}  
\affil{The Inter-University Institute for Data Intensive Astronomy (IDIA), Department of Astronomy,  and University of Cape Town,Private Bag X3, Rondebosch, 7701, South Africa, and University of the Western Cape, Department of Physics and Astronomy, Bellville, 7535, South Africa}
\affil{National Radio Astronomy Observatory, Socorro, NM 87801, USA}

\author{S. Sikhosana}  
\affil{Astrophysics Research Centre and School of Mathematics, Statistics and Computer Science, University of KwaZulu-Natal, Durban 4041, South Africa}

\author{J. Wagenveld}  
\affil{Max-Planck-Institut f\"ur Radioastronomie, Auf dem H\"ugel 69, D-53121 Bonn, Germany}

\author{O. I. Wong}  
\affil{CSIRO Astronomy and Space Science, PO Box 1130, Bentley, WA 6102, Australia}
\affil{ICRAR-M468, UWA, 35 Stirling Hwy, Crawley, WA 6009, Australia}
\affil{ARC Centre of Excellence for All Sky Astrophysics in 3 Dimensions (ASTRO 3D), Australia}

\begin{abstract}

We present details of the Automated Radio Telescope Imaging Pipeline ({\tt ARTIP}) and results of a sensitive blind search for \hi\ and OH absorbers at $z<0.4$ and $z<0.7$, respectively. {\tt ARTIP} is written in Python 3.6, extensively uses the Common Astronomy Software Application (CASA) tools and tasks, and is designed to enable the geographically-distributed  MeerKAT Absorption Line Survey (MALS) team to collaboratively process large volumes of radio interferometric data. We apply it to the first MALS dataset obtained using the 64-dish MeerKAT radio telescope and 32K channel mode of the correlator. With merely 40\,minutes on target, we present the most sensitive spectrum of PKS\,1830-211 ever obtained and characterize the known \hi\ ($z=0.19$) and OH ($z=0.89$) absorbers. We further demonstrate {\tt ARTIP}'s capabilities to handle realistic observing scenarios by applying it to a sample of 72 bright radio sources observed with the upgraded Giant Metrewave Radio Telescope (uGMRT) to blindly search for \hi\ and OH absorbers.  We estimate the numbers of \hi\ and OH absorbers per unit redshift  to be $n_{21}(z\sim0.18)<$0.14 and  $n_{\rm OH}(z\sim0.40)<$0.12, respectively, and constrain the cold gas covering factor of galaxies at large impact parameters (50\,kpc$<\rho<$150\,kpc) to be less than 0.022. Due to the small redshift path, $\Delta$$z\sim$13 for \hi\ with column density$>5.4\times10^{19}$\cmsq, the survey has probed only the outskirts of star-forming galaxies at $\rho>30$\,kpc. MALS with the expected $\Delta$$z\sim10^{3-4}$ will overcome this limitation and provide stringent constraints on the cold gas fraction of galaxies in diverse environments over $0<z<1.5$.

\end{abstract}

\keywords{quasars: absorption lines ---  interstellar medium}

\section{Introduction} 
\label{sec:intro}  
The neutral atomic hydrogen (\hi) 21-cm absorption line is an excellent tracer of the cold neutral medium (CNM) in the 
Galaxy \citep[e.g.][]{heiles03}. The CNM in the Galaxy is expected to have kinetic temperatures and volume densities 
in the range of $T_{\rm K}\sim$40-200\,K and log$\,n$(cm$^{-3}$)$\sim$1-2, respectively \citep[][]{Wolfire03}. 
Similarly, the OH radical which is a common constituent of diffuse (log$\,n$(cm$^{-3}$) $\sim$ 2) 
and dense (log$\,n$(cm$^{-3}$) $>$ 4) molecular clouds, 
can be observed through its 18-cm lines, namely, the main lines at 1665\,MHz and 1667\,MHz, and 
the two satellite lines at 1612\,MHz and 1720\,MHz \citep[e.g.,][]{Li18}.

The volume filling factor of various gas phases in the 
interstellar medium (ISM) and the observables associated with the \hi\ and OH absorption lines depend sensitively on in situ star 
formation through various stellar feedback processes \citep[][]{Liszt96, Wolfire03}. Therefore, observations of these lines 
can be used to determine physical conditions in the cold phases of the ISM and how they impact ongoing and future star 
formation in galaxies. In addition, the observed frequencies of the \hi\ and OH lines can be used to place the most stringent 
constraints on fractional variations of dimensionless fundamental constants of physics \citep[][]{Darling04, Kanekar05, Rahmani12, Gupta18oh} and 
cosmic acceleration \citep[][]{Darling12}.

Motivated by the scientific potential of these absorption lines, there have been tremendous efforts from the community over 
the past three decades to search for \hi\ and OH absorbers at $0 < z < 5$. But mainly due to the technical limitations imposed 
by narrow bandwidths and a hostile radio frequency environment,  majority of efforts at radio wavelengths so far has been to search for \hi\ 21-cm 
absorption in environments with higher chances of detection.  These include quasar sight 
lines with a damped Ly$\alpha$ absorber (DLA; with $N$(\hi)$\ge2\times10^{20}$\,\cmsq), a strong \mgii\ absorption line or a galaxy at small impact 
parameter ($\rho <$30\,kpc). 
However, these preselection methods based on optical and ultraviolet spectroscopic surveys are generally biased against sight lines through dense molecular and translucent 
phases of the ISM.  

While some individual measurements at optical and ultraviolet wavelengths indicate that cold atomic and molecular gas can 
be detected in low column density \hi\ absorbers (sub-DLAs; $10^{19} \le$ $N$(\hi) $< 2 \times 10^{20}$\,\cmsq) or at galaxy impact parameters as large as 50-80\,kpc 
\citep[][]{Muzahid15, Neeleman17, zou18}, statistical studies of H$_2$-bearing DLAs at $z\sim3$ show that the incidence rate (or cross-section) 
of the cold phase probed by H$_2$ absorption is $\sim$25 times less than that of \hi\ \citep[][see also \citealt{Zwaan06}]{Balashev18}. Additionally the incidence rate of cold gas 
is significantly enhanced in extremely strong DLAs ($N$(\hi) $\ge$ 5$\times 10^{21}$\,\cmsq), which generally probe galaxies at small impact parameters 
\citep[][]{Noterdaeme14, Ranjan18, Ranjan20}. 
The latter result also indicates that the cold gas is preferentially located at small impact parameters, i.e., less than a few kpc. 
For these impact parameters the dust bias can be very strong for optically selected quasars, leaving sight lines through the dense molecular and translucent phases of the 
ISM unexplored \citep[][]{Krogager19}. 

In coming years, these limitations will be overcome by large blind \hi\ and OH absorption line surveys planned with upcoming Square Kilometre Array (SKA) pathfinders and precursors. 
These include the MeerKAT Absorption Line Survey \citep[MALS;][]{Gupta17mals} using MeerKAT \citep[][]{Jonas16, Camilo18, Mauch20}, 
the First Large Absorption Survey in \hi\ \citep[FLASH;][]{Allison17} using ASKAP \citep[][]{Johnston07}, 
and the Search for \hi\ Absorption with AperTIF \citep[SHARP;][]{Adams18}. 

Along with these large surveys come large volumes of data and the associated Big Data challenges for which new techniques, 
algorithms and infrastructure need to be developed. 
Specifically,  MALS  has been allocated  $\sim$1655\,hours of MeerKAT observations using the L- and UHF-bands covering 900-1670\,MHz and 
580-1015\,MHz, respectively.
For $\sim$1100 MALS pointings, the radio interferometric data from MeerKAT 
will be acquired in both parallel- and cross-parallel hands using 32768 frequency channels. 
With this setup for a correlator 
dump time of 8\,seconds each survey pointing will generate about 1\,TB of data per hour, which for the total observing time  
will lead to a total raw visibility data volume of 1.7\,PB. These data cannot be processed using traditional serialized 
approaches. We have therefore been developing a data analysis pipeline called the {\tt Automated Radio Telescope Imaging Pipeline 
({\tt ARTIP})} in Python using the {\tt Common Astronomy Software Applications (CASA}) based tasks and tools. 
The main 
objective of the pipeline is to process the MALS data to produce final full-Stokes radio continuum images and 
spectral line cubes, with as little manual intervention as possible.  Another objective is to enable the geographically-distributed 
MALS team  to collaboratively set up, trigger, and monitor data processing and archiving in a centralized facility.

Although developed for MALS, the design of {\tt ARTIP} is general enough to accommodate datasets from any of the currently operational 
radio interferometers such as the Karl G. Jansky Very Large Array (VLA) and the upgraded Giant Metrewave Radio Telescope (uGMRT). 
This generalization is important for {\tt ARTIP}, because many algorithmic choices in MALS processing still require 
significant research, and are under active optimization and testing.  The flexibility enables {\tt ARTIP} to be tested on similar 
datasets from other telescopes well before the start of proper science observations of MALS.  

In this paper, we provide an overview of {\tt ARTIP} with a focus on the spectral line processing of wideband datasets. The processing 
of narrowband datasets has been demonstrated in \citet[][]{Gupta18oh, Gupta18j1243}.
The aspects of radio continuum and polarization processing will be presented in a future paper. 
Here, we validate the latest version of {\tt ARTIP} based on {\tt CASA 5.4-5.6} and its wideband spectral-line processing capabilities 
by applying it to a dataset ($\sim$1.5\,TB) having full L-band coverage obtained with the full MeerKAT 64-dish (MeerKAT-64) array. 
Then, we apply {\tt ARTIP} to a sample of bright 
radio sources observed using the uGMRT L-band receiver to carry out a blind search for \hi\ absorption at $0<z<0.4$ and OH absorption 
at $0.14<z<0.67$.  
The application to the uGMRT survey allows us to verify {\tt ARTIP}'s ability to handle datasets obtained under a broader range of 
realistic observing scenarios involving multiple calibrators and target sources. 
The integrated redshift path lengths \citep[$\Delta z$;][]{Lanzetta87} over which the absorbers with $N$(\hi)$>5\times10^{19}$\cmsq\ (spin temperature = 100\,K) and 
$N$(OH)$>2.4\times10^{14}$\,\cmsq\ (excitation temperature = 3.5\,K) can be detected in the uGMRT survey are 12.9 and 15.1, respectively. 
Based on these results, we present the most stringent constraints on the numbers per unit redshift range of \hi\ and OH absorbers.

This paper is laid out as follows.  In Section~\ref{sec:artip}, we present a brief overview of {\tt ARTIP} which was used to process the data presented in this paper. The readers interested in details may refer to Appendix~\ref{sec:artipdetails}. In Section~\ref{sec:pks}, we apply {\tt ARTIP} to the MeerKAT science verification observations of PKS\,1830-211, 
a well-known gravitationally lensed quasar at $z=2.507$. 
This quasar sight line is  unique: it has two \hi\ 21-cm absorbers, one at $z=0.190$ and another at $z=0.885$
\citep[][]{Lovell96, Wiklind96, Chengalur99}.
The details of the uGMRT survey, i.e., sample, observations, data processing, and absorption line analysis are presented in Section~\ref{sec:pilot}.
The results from the uGMRT survey are presented in Section~\ref{sec:res}.  In this section, we compute the incidences, i.e., numbers per 
unit redshift of \hi\ and OH absorbers, 
and compare these with the expectations from searches based on \mgii\ absorption and nearby galaxies. 
For completeness, the constraints on cold gas associated with the active galactic nuclei (AGNs) are also presented.
A summary of the results with future prospects is provided in Section~\ref{sec:summ}.

Throughout this paper we use the $\Lambda$CDM cosmology with $\Omega_m$=0.27, $\Omega_\Lambda$=0.73 and 
H$_{\rm o}$=71\,\kms\,Mpc$^{-1}$.

\section{ARTIP - Automated Radio Telescope Imaging Pipeline}      
\label{sec:artip}   
%

\begin{figure}
	\centerline{
\includegraphics[trim = {0.0cm 2.5cm 0.0cm 0cm}, width=0.35\textwidth,angle=0]{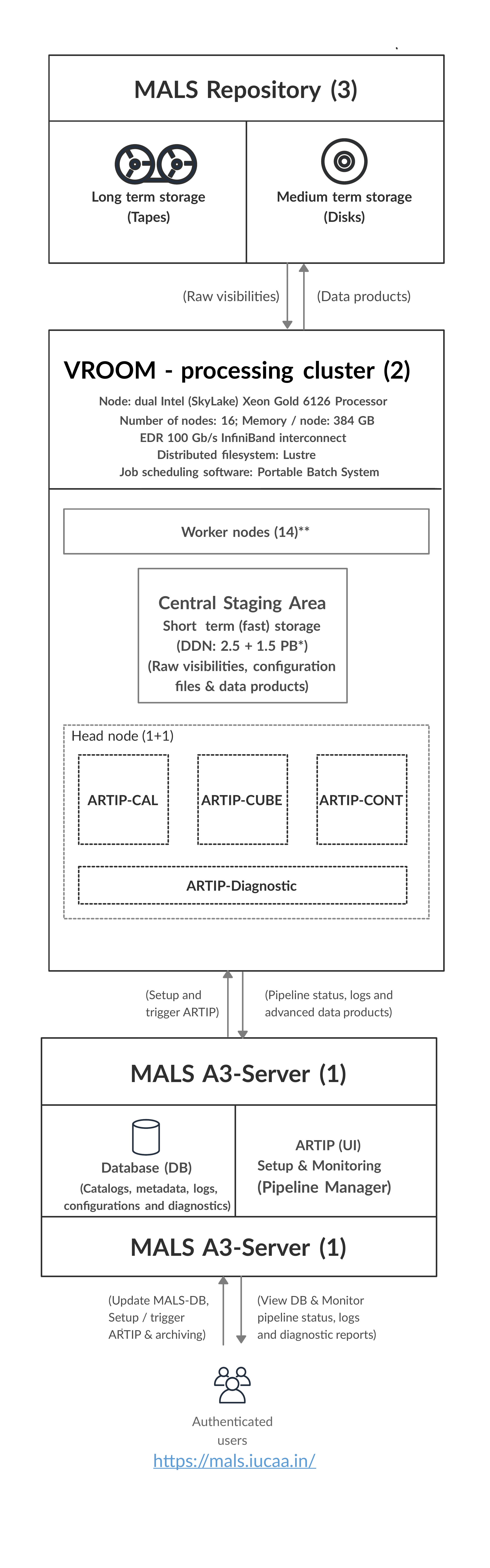} 
}
\vskip+0.0cm  
	\caption{ Schematic of {\tt ARTIP} deployment in the MALS A3, i.e., {\it Automated processing, Analysis, and Archiving} environment consisting 
	of (1) the A3-server, 
	(2) the VROOM-processing cluster, 
	and (3) the Repository.
	}
	\footnotesize{$^*$ VROOM is set up to support fast DDN storage of 4\,PB; currently 2.5\,PB are available.
	$^{**}$ The number of worker nodes is expandable through an interface with the general purpose cluster Pegasus with 64 nodes.}
\label{fig:artipscm}   
\end{figure} 

{\tt ARTIP} is being developed to perform end-to-end data processing for MALS. Besides carrying out a sensitive search for intervening \hi\ and OH 
absorbers, MALS will also provide an  unbiased  census  of intrinsic \hi\ and OH  absorbers,  i.e., cold  gas associated
with  powerful  AGNs (radio power $>10^{24}$\,W\,Hz$^{-1}$) at $0<z<2$, and will simultaneously deliver a blind \hi\ and OH emission 
line survey, and a full-polarization radio 
continuum survey.  The science goals as summarized in \citet[][]{Gupta17mals} are far reaching, and require the following {\tt data products}: 
{\it (i)} calibrated visibilities, {\it (ii)} wideband full primary beam Stokes I, Q, U and V images, and {\it (iii)} radio continuum subtracted 
Stokes I spectral line cubes at full spectral resolution. 
Therefore, {\tt ARTIP} is being designed with an extremely broad range of scientific use cases in mind.  

Overall, the outputs from the pipeline, hereafter referred to as {\tt pipeline products}, will be 
{\it (i)} {\tt data products}, as mentioned above, which can be used to generate absorption line catalogs, \hi\ moment maps, etc.; 
{\it (ii)} pipeline status and logs, which can be used to monitor and debug the processing; 
and {\it (iii)} diagnostic plots and statistics, which can be used to assess the quality of flagging, calibration, and data products.
These outputs will act as proxies to the raw visibility data, and the users concerned with specific science cases will not need to interact with the visibility data.

In general, the radio interferometric data prior to any imaging must be calibrated to recover the true or uncorrupted visibilities.  For an antenna pair, i.e., baseline $i-j$, 
the true visibility vector ($\overrightarrow{V}_{ij}^{T}$) representing four correlation combinations is related to the observed visibilities 
($\overrightarrow{V}_{ij}^{O}$) through the Hamaker-Bregman-Sault Measurement Equation \citep[][]{Hamaker96}
\begin{equation}
	\overrightarrow{V}_{ij}^{O} = M_{ij}\overrightarrow{V}_{ij}^{T}.
\label{eqhbs}
\end{equation}
 Here $M_{ij}$ = $J_{i} \otimes J_{j}^{*}$ is a 4$\times$4 Mueller matrix which contains the cumulative effect of all the measurement corruptions.  $J_{i}$ and $J_{j}$ encode a sequence of finite number of antenna-based corruptions (called  antenna gains). {\tt ARTIP}'s calibration and imaging sequence is essentially based on correcting for these corrupting effects utilizing the framework 
implemented in {\tt CASA}.  

{\tt ARTIP}  follows a stage driven architecture, in which outputs from previous stages are used by subsequent stages.    
At the highest level, it is split into the following four components:  
{\tt ARTIP-CAL}, {\tt ARTIP-CUBE}, {\tt ARTIP-CONT} and {\tt ARTIP-Diagnostic} which are responsible for calibration, spectral line imaging, continuum imaging, and generating various diagnostic plots and statistics, respectively.
It also consists of the {\tt Pipeline Manager} which offers a highly user-friendly interface to geographically-distributed MALS team to collaboratively 
{\it (i)} set up and trigger {\tt ARTIP} execution, monitor progress, and manage data products, and 
{\it (ii)} seamlessly chain together the outputs from {\tt CAL} that can be used by {\tt CUBE}, {\tt CONT}, and {\tt Diagnostic}.

In Fig.~\ref{fig:artipscm}, we present the complete deployment of {\tt ARTIP} in the MALS A3, i.e., {\it Automated Processing, Analysis, and Archiving} environment, which was used to process the data presented in this paper.
The specifications of {\tt VROOM}, the MALS processing cluster, i.e., {\tt VROOM} at the Inter-University Centre for Astronomy and 
Astrophysics (IUCAA) in India, are also provided in Fig.~\ref{fig:artipscm}.
Further details of {\tt ARTIP} are presented in Appendix~\ref{sec:artipdetails}. 
Specifically, the details of design, capabilities and deployment are provided in Appendix~\ref{sec:artipdesign}, and the processing stages are described in Appendix~\ref{sec:artipstages}.

%
\section{First MALS observation: PKS\,1830-211}      
\label{sec:pks}

\begin{figure*}
\includegraphics[trim = {0cm 1.0cm 0.0cm 0.0cm}, width=0.93\textwidth,angle=0]{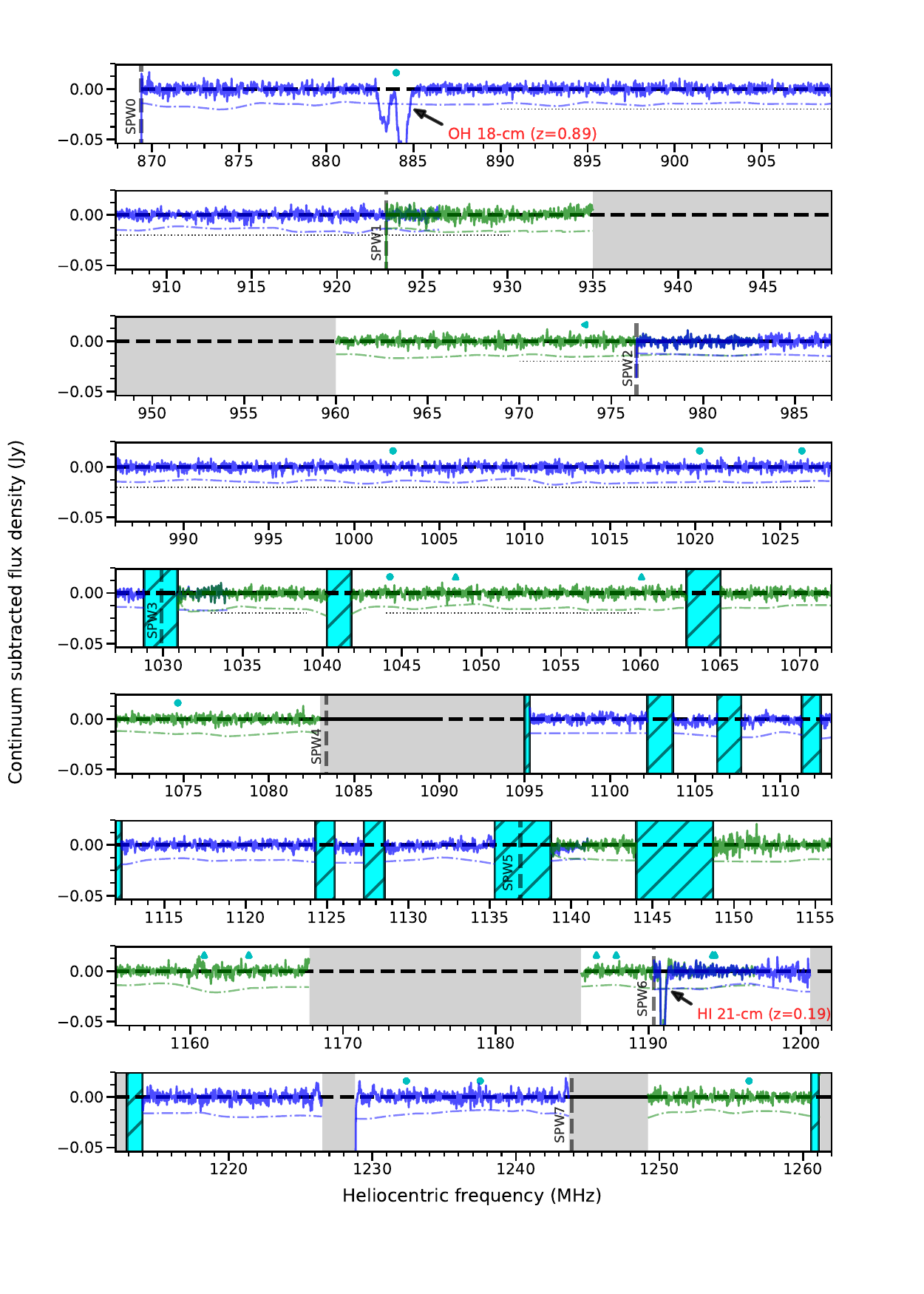}  
\vskip+0.0cm  
\caption{
	The continuum subtracted Stokes-$I$ spectrum of PKS\,1830-211. The adjacent spectral windows are plotted alternately in blue and green. 
	Shaded regions mark frequency ranges that were masked prior to any calibration.  
	Hatched regions were masked after calibration and imaging to exclude persistent RFI identified during the absorption line analysis.  
	Dash-dotted line is the detection threshold (5$\times\sigma_{\rm rolling}$).  The dotted line marks the frequency range used for the continuum imaging. 
	The redshifted frequencies ($z=0.88582$) of various NS, CCS, and CH$_2$CN transitions are marked using {\large $\circ$}, {\large $\triangleleft$}, and $\triangle$.
} 
\label{fig:j1830spec}   
\end{figure*} 

\begin{figure*}
\centering
\includegraphics[trim = {0cm 2.5cm 0.3cm 1.2cm}, width=1.0\textwidth,angle=0]{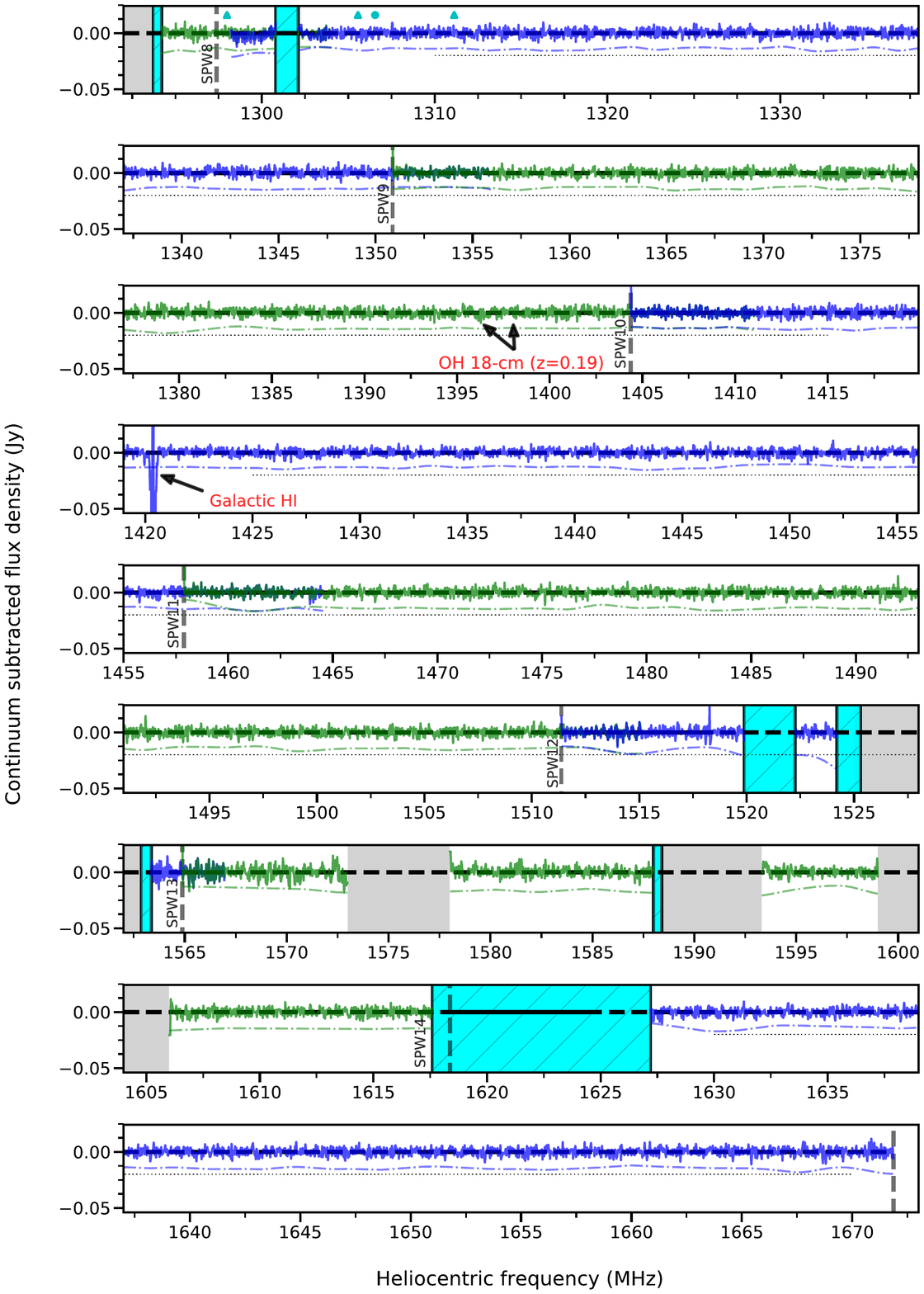}  
\\
	{\bf Figure \ref{fig:j1830spec}} {\it continued}.
\vskip+0.0cm  
\end{figure*} 

The field centered on PKS\,1830-211 was observed on December 19, 2019 with the MeerKAT-64 array. 
For the observation, the total bandwidth of 856\,MHz centered at 1283.9869 MHz split into 32768 frequency channels was used. 
This correlator mode delivers a frequency resolution of 26.123\,kHz, which is 6.1\,\kms\ at the center of the band. 
The correlator dump time was 8\,seconds.  For dual linearly  polarized L-band feeds with orthogonal polarizations labelled X and Y, the data were acquired for all four polarization products i.e., XX, XY, YX and YY. This is the first MALS 
observation using the SKA Reconfigurable Application Board (SKARAB) correlator's 32K mode. Of the 64 antennas, 63 participated 
in the observation; m052 was unavailable. 
At the start of observation, we observed PKS\,1934-638 for 15\,minutes for flux density, delay and bandpass calibrations. The target source was then observed 
for  40\,minutes. Since PKS\,1830-211 is a bonafide complex gain calibrator for the VLA in C- and D- array configurations, there was no need to 
separately observe a complex gain calibrator.
The full dataset in measurement set format is about 1.5\,TB, and was processed on the VROOM cluster using the latest version of {\tt ARTIP} based 
on CASA 5.6.1 

\subsection{Calibration}
\label{sec:pkscal}

Here, we are interested only in the Stokes-$I$ properties; therefore, for processing we generated a measurement set consisting of only XX and YY polarization  
products. We also dropped extreme edge channels.  The resultant measurement set with 30720 frequency channels was calibrated by processing data through stages (1) to (7) of {\tt ARTIP-CAL} (see Appendix~\ref{sec:artipstages} for details). 

In the first stage of processing, an initial RFI mask to eliminate the strongest RFI spikes that were present 
even on the longest baselines was applied (see shaded regions in Fig.~\ref{fig:j1830spec}).
After this, the model visibilities for the calibrator defined by {\tt Stevens-Reynolds 2016} flux density scale 
([I=13.993, Q=0, U=0, V=0] Jy @ 8.56e+08Hz) 
were predicted, and the pipeline performed initial calibration on a subset of frequency channels (19000$-$20000) to identify any non-working antennas or 
baselines. No additional flagging was needed at this stage.

Next, the pipeline proceeded to calibrate the entire band. First, strong RFI was flagged using {\tt tfcrop}. Then, flux density, delay, and 
bandpass calibration were performed. After applying these calibrations but without any calibration-based flagging, the data on PKS\,1934-638 were 
further flagged for RFI using {\tt rflag}. The flux density, delay, and bandpass calibrations were then performed again. All the delays were in the 
range (-0.4, 0.4) nanoseconds, and we did not notice any change in these before or after {\tt rflag}; however, bandpass solutions after the RFI flagging 
were substantially improved.

Then, the pipeline derived the gain calibration solutions on PKS\,1830-211 and scaled the amplitude of the solutions with respect to the known flux density calibrator.  
Finally, in the last processing stage, all the calibrations were applied to the target source visibilities. 

The calibrated data were then processed using {\tt ARTIP-CONT} and {\tt ARTIP-CUBE} to perform wideband continuum and spectral line imaging, respectively.

\subsection{Continuum imaging}
\label{sec:pkscont}

Briefly, the data were first averaged in frequency per 32 channels 
($\sim$0.8\,MHz) and a more stringent RFI mask to completely exclude band edges and RFI-afflicted regions was applied. The frequency range considered 
for continuum imaging is shown with horizontal dotted lines in Fig.~\ref{fig:j1830spec}.  Additionally, the resultant frequency-averaged 960 channels 
were regridded along the frequency axis to obtain a measurement set with 16 physically distinct spectral windows.
We created a widefield broad band 6k$\times$6k continuum image with a pixel size of $2^{\prime\prime}$, spanning $\sim3.3^\circ$, using {\tt tclean} in 
CASA. The {\tt w-projection} algorithm was used as the gridding algorithm in combination with {\tt Multi-scale Multi-term Multi-frequency synthesis} 
({\tt MTMFS}) for deconvolution, with nterms = 2 and four pixel scales to model the extended emission \citep[see][]{Rau11, Bhatnagar13, Jagannathan17}. 
Two rounds of phase-only self-calibration were carried out, along with a final round of amplitude and phase self-calibration. Imaging masks were 
appropriately adjusted between major cycles during imaging and self-calibration.  The objective of developing {\tt CONT} is to automate this 
process.  The final continuum image made using {\tt robust=0} weighting 
has a synthesized beam of $12.9^{\prime\prime}\times8.1^{\prime\prime}$ (position angle = $-76.3^\circ$).  The rms is 140\,$\mu$Jy\,beam$^{-1}$ close to the 
bright radio source at the center and 40\,$\mu$Jy\,beam$^{-1}$ (dynamic range $\sim$78500) away from it.  The latter is about a factor of 4 higher than the theoretical thermal noise, 
most likely due to the dynamic range limitation (see also Section~\ref{sec:summ}). 

The quasar PKS\,1830-211 is known to be variable at radio wavelengths\footnote{Between 1996-2016, the flux density at $0.8 - 1.4$\,GHz varied between 
10-14\,Jy \citep[see Fig.~7 of][]{Allison17}.}.  
In the uv-plane, at 1.4\,GHz its 
radio emission is resolved beyond 16\,k$\lambda$ (see VLA calibrator manual\footnote{https://science.nrao.edu/facilities/vla/observing/callist}). 
A sub-arcsecond scale resolution image of this gravitationally lensed object obtained using the Multi-Element Radio Linked Interferometer Network (MERLIN) shows that the radio morphology consists of two compact components separated 
by 1$^{\prime\prime}$ and surrounded by a low-surface brightness elliptical Einstein ring \citep[][]{Rao88, Jauncey91}.

For comparison, the longest baseline length in our MeerKAT dataset is $\sim$30\,k$\lambda$.
The continuum flux density of the quasar in the wideband MTMFS image was recovered to be 11.245 $\pm$ 0.001\,Jy at the reference frequency 
of 1270\,MHz.  The in-band integrated spectral index $\alpha$ = 0.13.
As expected, the radio emission is only barely resolved in the MeerKAT image.  The quoted uncertainty on the flux density corresponds to errors from the single Gaussian component fitted to the continuum image.  The flux density accuracy at these low frequencies are limited by a variety of factors such as the accuracy of overall calibration, and is expected to be about $\sim$5\%. 

\subsection{Cube imaging}
\label{sec:pkscube}

We split 30720 frequency channels into 15 spectral windows (SPWs) labeled as SPW0 to SPW14.  The starting frequencies of these measurement sets are shown as 
vertical dashed lines in Fig.~\ref{fig:j1830spec}, and the adjacent SPWs are plotted alternately in blue and green.  Note that the adjacent SPWs 
have an overlap of 256 channels ($\sim$7 MHz).  The overlap ensures that no spectral features at the edge of any SPW are lost. 

The measurement sets for these SPWs are then processed for continuum imaging with self-calibration and cube imaging, i.e., stages-(2) to -(4) of {\tt ARTIP-CUBE}. 
For continuum imaging using {\tt w-projection} and {\tt MTMFS} in {\tt ARTIP-CUBE}, a continuum dataset is generated by 
flagging RFI-affected frequency ranges and averaging data in frequency by 32 channels to reduce the data volume. 
Self-calibration is initiated by predicting model visibilities 
based on the wideband continuum image obtained through {\tt ARTIP-CONT}. Here also two rounds of phase-only and one round of amplitude-and-phase 
self-calibration were performed.

The self-calibration solutions were then applied to the line dataset. The continuum subtraction was performed using the image, i.e., CLEAN 
components obtained from the last round of self-calibration. The continuum-subtracted visibilities were then inverted to obtain 
XX and YY spectral line cubes with {\tt robust=0} weighting and 128$\times$128 pixels.

\subsection{Spectral line analysis}
\label{sec:pkscube}

For absorption line identification and analysis, the XX and YY spectral line cubes (SPW-cubes) were smoothed to a common resolution of 
$21.0^{\prime\prime}\times13.9^{\prime\prime}$ (position angle = $-72.0^\circ$)\footnote {Synthesized beam of the SPW0-cube, whose central frequency is 899.4296\,MHz.} and the spectra were extracted for all SPW-cubes at the 
location of PKS\,1830-211. 
The  residual flux density in each spectrum is of the order of 0.5\% of the flux density of PKS\,1830-211.  The shape of the residual flux is generally 
anti-correlated with respect to the spectrum of PKS\,1934-638, which peaks (turnover) at $\sim$1220 MHz. 

The final XX and YY spectra were obtained by removing the residual continuum by fitting a low-order polynomial.  The procedure involved 
iteratively clipping deviant pixels to find a smooth continuum fit unaffected by the presence of various spectral line features.
After this step, an error spectrum was estimated by calculating the rolling standard deviation ($\sigma_{\rm rolling}$) with a window size of 32 channels.
In this step the deviant pixels were also iteratively clipped, and the error spectrum was interpolated across the masked region.
We searched these spectra for  emission and absorption lines using a detection 
algorithm that requires at a channel $j$, (i) flux density $\mid$F($j$)$\mid$ $>$$5\times\sigma_{rolling}$($j$),  
and (ii) heliocentric frequency $\nu$($j$) $\ge \nu_o$/(1 + $z_{PKS1830-211}$), where $\nu_o$ is the rest-frame \hi\ 21-cm or OH 18-cm line frequency. 
For an absorption candidate to be deemed real, it must be consistently (difference$<3\sigma$) detected in both the XX and YY spectra.  
The hatched regions in Fig.~\ref{fig:j1830spec} are frequency ranges with spurious (i.e., due to statistical fluctuation or RFI) 
spectral line features identified through the detection algorithm.  These candidates can then be grouped and listed for visual inspection and 
various statistical tests.  Further details  will be provided in a future paper (Gupta et al. in preparation).

We confirm the detection of previously known \hi\ 21-cm absorption ($z=0.19$) and OH 18-cm main line absorption ($z=0.89$) with extremely high significance 
(signal-to-noise ratio $\sim$70),
but except for Galactic \hi, we do not detect any other true absorption.   
The final unsmoothed Stokes-I spectrum of PKS\,1830-211 obtained by averaging XX and YY spectra with 
appropriate weights is presented in Fig.~\ref{fig:j1830spec}.  The dash-dotted line in the figure represents the detection threshold (5$\times\sigma_{\rm rolling}$). 
The spectral rms in the Stokes-$I$ spectrum at 1360\,MHz is 2.6\,mJy\,beam$^{-1}$ (1$\sigma$ for {\tt robust=0}). 
For $T_{\rm s}$ = 100\,K, $T_{\rm ex}$ = 3.5\,K, unity covering factor, and line width of 6\,\kms, this rms corresponds to a 5$\sigma$ sensitivities to detect gas with   
column densities of $N$(\hi) $>$ 1.4$\times10^{18}$\,\cmsq\ and $N$(OH) $>$ 5.9$\times10^{12}$\,\cmsq, respectively 
(see Equations~\ref{eq21cm} and \ref{eqoh}).  These limits are well below the column density requirements of MALS.   

\subsubsection{Absorption at $z=0.19$}
\label{sec:pkslowz}

\begin{figure}
\hbox{
\includegraphics[trim = {1.0cm 5.0cm 0.0cm 3.0cm}, width=0.50\textwidth,angle=0]{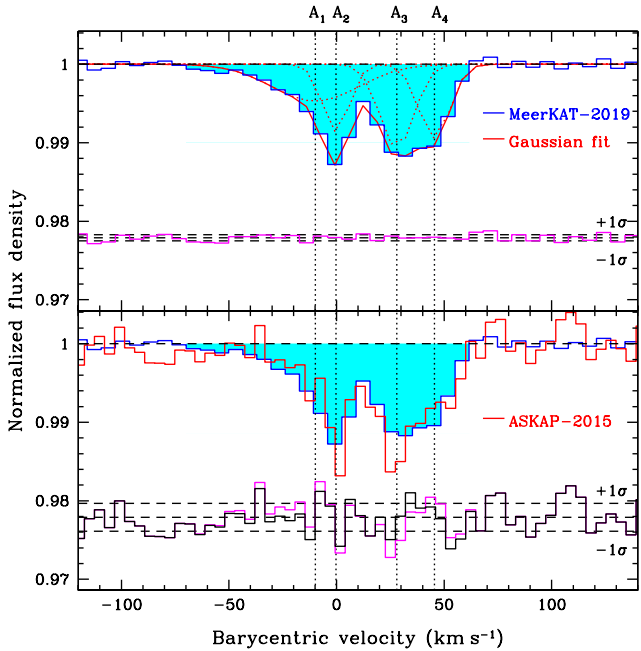}  
}
\caption{
	MeerKAT stokes-$I$ spectrum (blue) of \hi\ 21-cm absorption at $z=0.19$ towards PKS\,1830-211.  
The spectral rms  is 3.4\,mJy\,beam$^{-1}$\,channel$^{-1}$. 
The zero of the velocity scale is defined at $z$=0.19259, the \hi\ absorption peak. 
{\it Top:} individual Gaussian components (Table~\ref{tab:21cmfit}) and the resulting 
fit to the stokes-$I$ spectrum are overplotted  as dotted and continuous lines (red), respectively. 
Vertical dotted lines mark the location of  components.
The residuals, with an arbitrary offset for clarity, are also shown.   
{\it Bottom:} The ASKAP spectrum from \citet[][]{Allison17} is overplotted. 
	The difference spectrum ([ASKAP spectrum] - [fit to the MeerKAT spectrum]) is shown in magenta.
	The residuals in black correspond to a constrained fit to the ASKAP spectrum (see text for details).
}
\label{pks21cm}
\end{figure}

In Fig.~\ref{pks21cm}, we zoom in on the \hi\ 21-cm absorption at $z=0.19$.  The absorption was originally detected by 
\citet[][]{Lovell96} using the Australia Telescope Compact Array (ATCA).
The double-peaked 21-cm absorption profile is also evident in our MeerKAT spectrum.  
The XX and YY spectra are also consistent with each other within statistical uncertainties.
From the stokes-$I$ spectrum, we measure the integrated 21-cm optical depth, $\int \tau (v) dv = 0.760\,\pm\,0.010$\,\kms. 
90\% of the total 21-cm optical depth is contained within $\Delta$V$_{90}$ = 86 $\pm$ 8\,\kms.    

\begin{table}
\caption{Multiple Gaussian fits to the absorption lines.}
\vspace{-0.4cm}
\begin{center}
\begin{tabular}{cccc}
\hline
\hline

{\large \strut}     Id.    &    Centre     &  $\sigma$            &   $\tau_p$       \\
                           &   (\kms)      &   (\kms)             &   ($10^{-3}$)    \\
\hline
	\multicolumn{4}{c}{ \hi\ absorption ($z$=0.19259) } \\
\hline
                A$_1$      &  -10 $\pm$ 7       & 20 $\pm$ 3       &    4.7 $\pm$ 0.8       \\
                A$_2$      &    0 $\pm$ 1       &  6 $\pm$ 1       &    8.6 $\pm$ 1.5        \\
                A$_3$      &   28 $\pm$ 1       &  8 $\pm$ 1       &   10.7 $\pm$ 1.4        \\
                A$_4$      &   45 $\pm$ 1       &  7 $\pm$ 1       &    9.4 $\pm$ 1.0        \\
\hline
	\multicolumn{4}{c}{ OH 1667\,MHz absorption ($z=0.88582$) } \\
\hline
                A$_1$      &  -214 $\pm$ 5       & 25 $\pm$ 5       &    1.4 $\pm$ 0.3       \\
                A$_2$      &  -113 $\pm$ 4       & 48 $\pm$ 5       &    6.0 $\pm$ 0.1       \\
                A$_3$      &  -7 $\pm$ 5       & 39 $\pm$ 3       &    4.6 $\pm$ 0.4       \\
\hline
	\multicolumn{4}{c}{ OH 1665\,MHz absorption ($z=0.88582$) } \\
\hline
                B$_1$      &  139               & 25               &    0.6 $\pm$ 0.2       \\
                B$_2$      &  239               & 48               &    3.4 $\pm$ 0.1       \\
                B$_3$      &  345               & 39               &    2.4 $\pm$ 0.2       \\
\hline
\end{tabular}
\tablecomments{ 
The centers of OH components are on the velocity scale defined for the 1667\,MHz line.   
	}
\label{tab:21cmfit}
\end{center}
\end{table}

For an optically thin cloud the integrated 21-cm optical depth (${\cal{T}} \equiv \int\tau dv$) is related to the neutral hydrogen column density 
$N$(H~{\sc i}), spin temperature $T_{\rm s}$, and covering factor $f_c^{\tiny \hi}$ through
\begin{equation}
	N{(\hi)}=1.823\times10^{18}~{T_{\rm s}\over f_{\rm c}^{\tiny \hi}}\int~\tau(v)~{\rm d}v~{\rm cm^{-2}}.
\label{eq21cm}
\end{equation}
\citet[][]{Verheijen01} used the VLA with the Pie Town antenna of the Very Long Baseline Array (VLBA) to resolve the double image 
of the lensed radio core.
They extract 21-cm absorption spectra towards the North-East (NE; peak intensity\,$\sim$5\,Jy) and South-West (SW; peak intensity\,$\sim$5\,Jy) 
continuum peaks separated by $\sim$1$^{\prime\prime}$ (3.2\,kpc at $z=0.19$). 
It is clear from their spectra that the majority of the absorption at 0\,\kms\ in Fig.~\ref{pks21cm} is seen only towards the SW component, 
whereas the absorption at +35\,\kms\ is from the NE component.  
For the MeerKAT spectrum, this dissection implies that $f_{\rm c}^{\tiny \hi} = 0.5$ and  $N$(\hi) = $(2.77 \pm 0.04)\times10^{20}$( ${T_{\rm s}}/{100\,{\rm K}})({0.5}/{f_{\rm c}^{\tiny \hi}} $ )\,\cmsq.
In general, spectroscopic observations using Very Long Baseline Interferometry (VLBI) reveal optical depth variations on scales of 
10-100\,pc \citep[see Section 3.4 of][]{Gupta18j1243}. 
Therefore, it is quite likely that $f_{\rm c}^{\tiny \hi}$ is even smaller than 0.5,  implying that the $N$(\hi) estimated above should be taken 
as a lower limit.


No OH 18-cm main lines at 1396.38 and 1398.03\,MHz corresponding to the $z=0.19$ absorber are detected.  For the 1667\,MHz line, the 5$\sigma$ detection limit 
corresponds to $N$(OH) $<$ 1.2$\times10^{13}$(${T_{\rm ex}}/{3.5\,{\rm K}}$)(${1.0}/{f_c^{\rm OH}}$)\,\cmsq\ (see Equation~\ref{eqoh} in next section).
Here, $T_{\rm ex}$ = 3.5\,K is the peak of the log-normal function fitted to the $T_{ex}$ distribution of OH absorbers observed in the Galaxy \citep[][]{Li18}.

In the lower panel of Fig.~\ref{pks21cm}, we overplot the \hi\ 21-cm absorption profile from the ASKAP spectrum obtained in 2015 by \citet[][]{Allison17}.  
For a detailed comparison between the two spectra, we model the MeerKAT spectrum using multiple Gaussian components.  The overall structure of the 
absorption line is reasonably modeled by a four-component fit, which is summarized in Table~\ref{tab:21cmfit} and shown in the top panel of Fig.~\ref{pks21cm}. 
The same Gaussian components also provide an acceptable fit to the ASKAP spectrum (see residuals plotted in magenta in the bottom panel). 
The integrated optical depths match within 1$\sigma$ uncertainty.
To check the overall shift between the two profiles, we also fitted the ASKAP spectrum by tying the widths and relative separation of 
the components to the fit in Table~\ref{tab:21cmfit}.  This exercise leads to a marginally better fit to the ASKAP profile (see residuals in black); 
 the relative shift between the MeerKAT and the ASKAP profiles is $<$ 1.7\,\kms.

\subsubsection{Absorption at $z=0.89$}
\label{sec:pkshighz}

\begin{figure}
\hbox{
\includegraphics[trim = {1.0cm 5.0cm 0.0cm 3.0cm}, width=0.50\textwidth,angle=0]{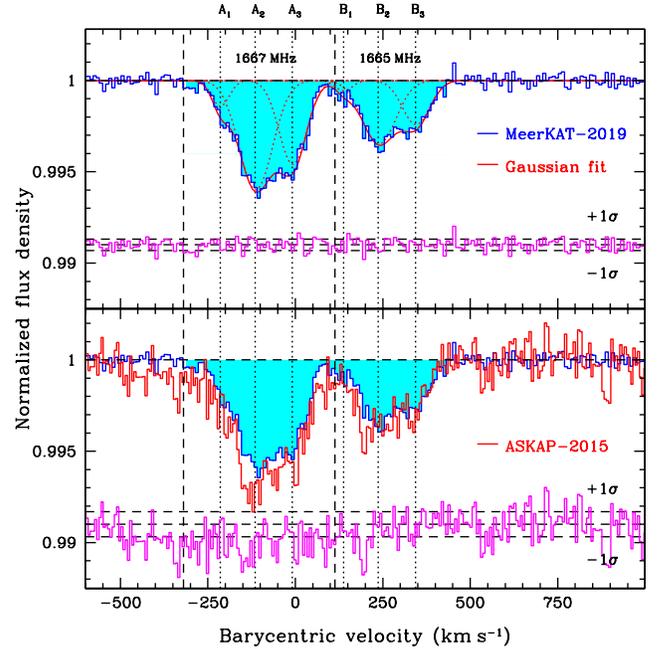}  
}
\caption{
	MeerKAT stokes-$I$ spectrum (blue) of OH absorption at $z=0.89$ towards PKS\,1830-211.  
The spectral rms  is 3.3\,mJy\,beam$^{-1}$\,channel$^{-1}$. 
	The zero of the velocity scale is defined for the 1667\,MHz line at $z$=0.88582, which corresponds to the main molecular (e.g., CO, HCO$^+$ and HCN) absorption 
	component detected at mm wavelength \citep[][]{Wiklind98}. 
	The vertical dashed lines mark the range over which \hi\ 21-cm absorption is detected \citep[][]{Chengalur99, Allison17}.
{\it Top:} individual Gaussian components (Table~\ref{tab:21cmfit}) and the resulting 
fit to the stokes-$I$ spectrum are overplotted  as dotted and continuous lines (red), respectively. 
Vertical dotted lines mark the locations of  components.
The residuals, arbitrarily offset for clarity, are also shown.   
{\it Bottom:} The ASKAP spectrum from \citet[][]{Allison17} is overplotted. 
	The difference spectrum ([ASKAP spectrum] - [fit to the MeerKAT spectrum]) is shown in magenta.
}
\label{fig:pksoh}
\end{figure}

In Fig.~\ref{fig:pksoh}, we zoom in on the OH absorption at $z=0.89$. This feature was originally detected by \citet[][]{Chengalur99}. 
Both the 1665 and 1667\,MHz lines are detected in the MeerKAT spectrum, and the absorption is confined to the velocity range over 
which the \hi\ 21-cm absorption is detected (see vertical dashed lines in Fig.~\ref{fig:pksoh}).  
Although only $\sim$30\,MHz from the edge of the L-band and affected by the bandpass rolloff, the spectrum exhibits exquisite sensitivity.
The integrated optical depths of the two lines as measured from the stokes-$I$ spectrum are 0.688\,$\pm$\,0.018\,$\pm\,0.071_{sys}$ 
and 1.265 $\pm$ 0.018\,$\pm\,0.071_{sys}$\,\kms, respectively.  
Here, 0.018 is the statistical error and $0.071_{sys}$ represents the systematic error due to the continuum placement 
uncertainties, quantified by fitting the XX and YY spectra independently with polynomials of different orders.
The optical depths are remarkably consistent with the 5:9 ratio expected for the two main lines originating from the gas in local thermodynamic equilibrium (LTE).  
From the stronger (i.e., the 1667\,MHz) line, we measure $\Delta V_{90}$ = 248 $\pm$ 9\,\kms. 

For an optically thin cloud, the integrated OH optical depth of the 1667\,MHz line is related to the OH column density $N$(OH) through
\begin{equation}
N{({\rm OH})}=2.24\times10^{14}~{T_{\rm ex}\over f_{\rm c}^{\rm OH}}\int~\tau_{1667}(v)~{\rm d}v~{\rm cm^{-2}}, 
\label{eqoh}
\end{equation}
where $T_{\rm ex}$ is the excitation temperature in Kelvin, $\tau_{1667}$($v$) is the optical depth of the 1667\,MHz line at 
velocity $v$, and $f_c^{\rm OH}$ is the covering factor \citep[e.g.][]{Liszt96}.
If we adopt 
$T_{\rm ex}$ = 5.14\,K, i.e., coupled to the cosmic microwave background (CMB), $T_{\rm CMB}$ at $z=0.89$,
we estimate $N$(OH) = (1.46  $\pm$ 0.05)$\times$10$^{15}$(${T_{\rm ex}}/{5.14\,{\rm K}}$)(${1.0}/{f_c^{\rm OH}}$)\,\cmsq.  
This value is similar to the column densities observed in dense molecular clouds in the Galaxy. 

The $z=0.89$ absorber is particularly special: at high frequencies numerous molecular species, such as CO, HCO$^+$, HCN, HNC, and their rare 
isotopologues have been detected towards the NE and SW continuum peaks \citep[e.g.,][]{Wiklind98, Muller14}. 
The absorber is also known to be rich in dust, with possibly crystalline silicate grains \citep[e.g.,][]{Aller12}. 
Spatially resolved observations of the two lensed components reveal that the sight line toward the SW component, 
which passes through a spiral arm of the lensing galaxy, is dustier and more molecule-rich. 
The compactness of the radio emission at high frequencies and the absorption line variability implies that the molecular clouds responsible for the absorption have characteristic sizes of 0.5-1\,pc \citep[][]{Muller08, Schulz15}. 
This result implies $f_c^{\rm OH} \ll $1 and most likely $f_c^{\rm OH}$ $\le$ $f_{\rm c}^{\tiny \hi}$. The $N$(OH) estimated above is strictly a lower limit.  

At high frequencies, the strongest molecular absorption lines are detected at 0\,\kms\ over a range of $\sim$100\,\kms. These are mostly 
associated with the SW component, which is closer to the center of the lensing galaxy and more obscured.  The sight line towards the NE component 
passes farther from the center of the lensing galaxy and exhibits much weaker well-detached absorption at about -150\,\kms.
The OH absorption is clearly detected over this entire velocity range, and compared to the high-frequency molecular lines, is more widespread 
in velocity space (from -350 to 100\,\kms). 

For a comparison with high frequency molecular lines, we model the two OH lines using multiple Gaussian components.  We assume that both the main lines 
originate from the same gas. Therefore, the centers and widths of the three components, A$_1$, A$_2$ and A$_3$, fitted to the 1667 MHz line 
are tied to those of components (B$_1$ - B$_3$) fitted to the 1665 MHz line (see Table~\ref{tab:21cmfit}).     
The components and the resultant fit are plotted in the upper panel of Fig.~\ref{fig:pksoh}.
We note that the ratio 5:9 expected for LTE also holds for the individual components.
Interestingly, the strongest OH absorption (component A$_2$)  corresponds to the velocities where little or no molecular absorption is 
detected at high frequencies.  Such differences could either be due to the different radio morphologies at low and high-frequency or imply that 
a part of the absorbing gas is CO-dark.  
Sub-arcsecond scale spectroscopy with low-frequency receivers is required to distinguish between these possibilities \citep[][]{paragi15}.

The MeerKAT spectrum has a total bandwidth of 856\,MHz and  covers several transitions of NS, CCS, and CH$_2$CN for the $z=0.89$ absorber.  
The redshifted frequencies of transitions unaffected by RFI are marked in Fig.~\ref{fig:j1830spec}.  None of these are detected.  
Assuming that the excitation is coupled to the CMB ($T_{\rm CMB}$=5.14\,K), $f_c$ = 1, and considering only the strongest transitions, we estimate 
5$\sigma$ upper limits on the column densities of NS, CCS and CH$_2$CN to be $9\times10^{14}$, $6\times10^{13}$ and $8\times10^{14}$\,\cmsq, respectively.

Finally, in the lower panel of Fig.~\ref{fig:pksoh}, we overplot the ASKAP absorption profile from \citet[][]{Allison17}. 
It is apparent from the profiles and the difference spectrum that while the MeerKAT and ASKAP profiles of the 1665\,MHz line match well 
over 350-550\,\kms, the absorption in the ASKAP spectrum is systematically larger at velocities $<$ 350\,\kms. 
Also, absorption in the ASKAP spectrum is detected at velocities ($>$305\,\kms) where no 21-cm absorption is detected
(refer to vertical dashed lines in Fig.~\ref{fig:pksoh}). The OH absorption at these velocities is also absent in the spectrum presented in \citet[][]{Chengalur99}.  
The excess in the ASKAP spectrum is caused by un-subtracted spectral structure in one epoch (2015 Oct 21) of commissioning observations  \citep[][private communication]{Allison17}.

\section{uGMRT absorption line survey}      
\label{sec:pilot}   

The successful processing of the PKS\,1830-211 dataset from MeerKAT has demonstrated the effectiveness of {\tt ARTIP} in streamlining the processing of a large dataset on a cluster.  
To further test the overall survey strategy involving target selection, absorption line identification, etc., and the end-to-end processing of a large number of datasets, 
we apply {\tt ARTIP} to a pilot absorption line survey using newly commissioned wideband observing modes of the uGMRT L-band system.

\subsection{Sample of radio sources}      
\label{sec:samp}   

\begin{figure} 
\centerline{\vbox{
\centerline{\hbox{ 
\includegraphics[trim = {0.0cm 0cm 1.5cm 0cm}, width=0.55\textwidth,angle=0]{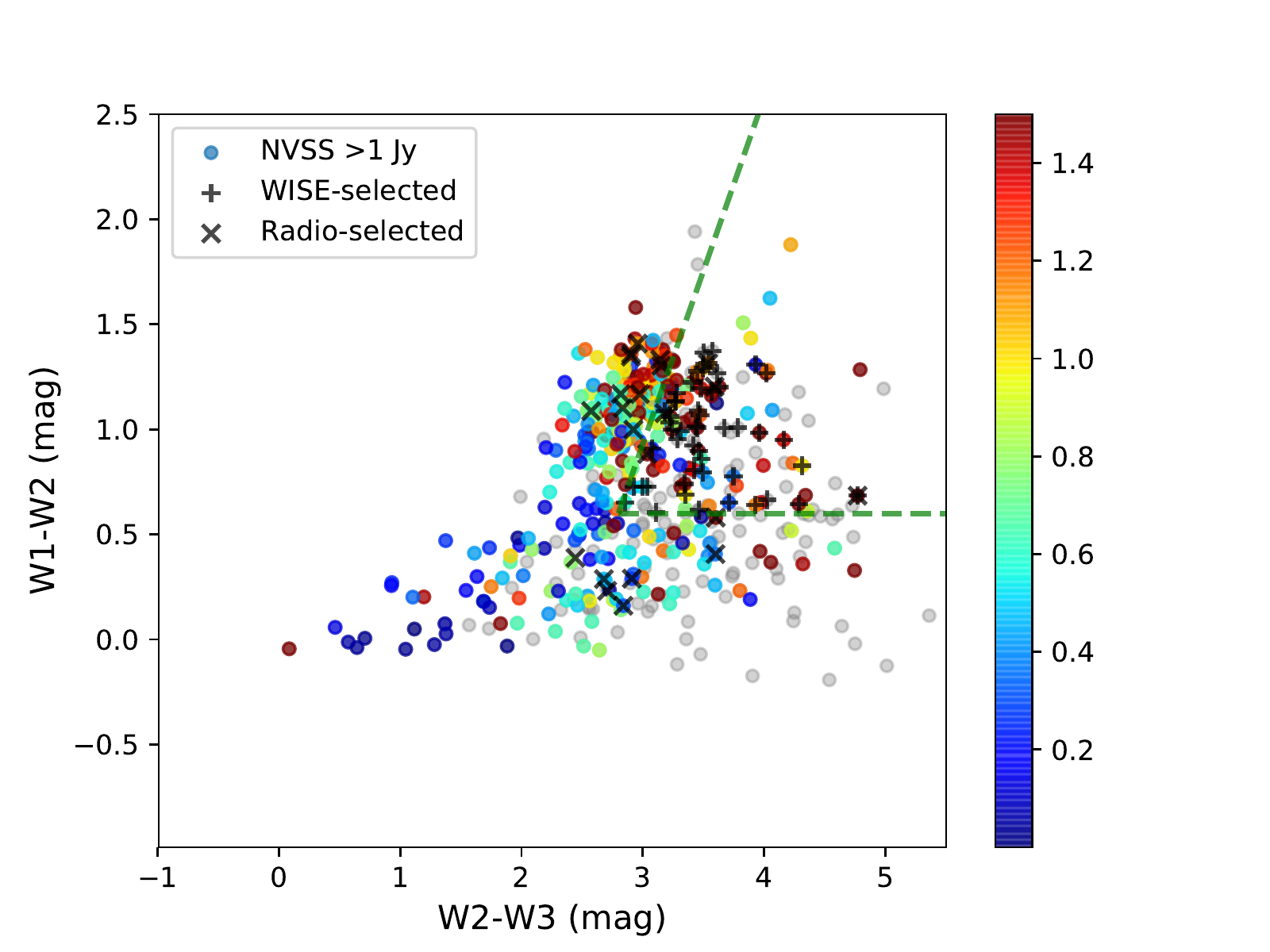}  
}} 
}}  
\vskip+0.0cm  
\caption{
	WISE infrared color-color plot. Radio sources brighter than 1\,Jy from NVSS are plotted as circles.  
	The objects with known redshift are color-coded as a function of redshift.  
The objects from WISE-selected and radio-selected samples are marked as $+$ and $\times$, respectively.  
	The dashed lines mark the region defined by Equation~\ref{eqwise}.
} 
\label{fig:wisecolor}   
\end{figure} 

We focus on selecting a dust-unbiased sample of the brightest radio loud quasars at $z>0.4$. 
The redshift cut-off is based on the lowest frequency (1000\,MHz) covered in the observations.   
We base our sample selection on two all-sky surveys: the  NVSS (resolution$\sim45^{\prime\prime}$) and the AllWISE catalogs.  
The former allows us to identify bright radio sources at 20\,cm, and the latter to select 
high-$z$ quasars. These two factors maximise the optical depth sensitivity and redshift path of the survey.     
Further, for this pilot survey we select targets that also have higher spatial resolution (5$^{\prime\prime}$) images from the Faint  Images  of the Radio  Sky  
at Twenty  centimetres  (FIRST) survey \citep[][]{Becker95} 
to ensure that the radio emission is indeed compact at scales of a few arcseconds.  We do not impose any right ascension or declination cuts.

In NVSS, there are 484 radio sources brighter than 1\,Jy that also {\it (i)} have a counterpart within $10^{\prime\prime}$ in the $W_1$ (3.4$\mu$m), $W_2$ (4.6$\mu$m) and $W_3$ (12$\mu$m) bands 
of the WISE survey, and {\it (ii)} have been observed in the FIRST survey. 
The ($W_1 - W_2$) $-$ ($W_2 - W_3$) color-color plot of these radio sources is shown in Fig.~\ref{fig:wisecolor}.  
The dashed lines in the figure correspond to following color cuts:
\begin{equation}
	\begin{array}{c}
W_1 - W_2 < 1.3\times(W_2 - W_3) - 3.04; \\
\\
W_1 - W_2 > 0.6.
\\
\\
\end{array}
\label{eqwise}
\end{equation}
They enclose the region that efficiently separates high-$z$ ($z>1.5$) quasars from radio galaxies and lower redshift quasars (Gupta et al. in preparation).  
The 117 radio sources satisfying these WISE color cuts form the 
parent sample for our survey.  The properties of 50 sources picked at random from the parent sample and observed as part of the pilot survey are provided in Table~\ref{tab:wisesamp}. 
Based on FIRST, 47/50 of these are compact at arcsecond scales. J0854+1405, J1011+4628 and J1612+2222 are extended in FIRST but with peak flux densities $>$1\,Jy.  The spectroscopic 
redshifts are known for 34/50 sources \citep[see column 4 of Table~\ref{tab:wisesamp} and ][]{Krogager18} and, as expected, only 10\% are at $z<0.4$.  

We augment the above sample with 22 radio sources that were observed in the uGMRT survey  as gain calibrators.  The radio and infrared 
properties of these sources are summarized in the later part of Table~\ref{tab:wisesamp}. As expected, all are compact on few arcsecond scales.
Only five sources (J0805+6144, J1150-0023, J1227+3635, J1405+0415 and J1640+1220) satisfy the above-mentioned WISE color cuts, but spectroscopic redshifts 
are available for 21/22 objects and only 4/21 with the confirmed redshifts are at $z<0.4$.
We emphasize that the addition of this subset of 22 sources selected purely on the basis of their radio properties does not add any dust bias to 
our overall sample.

In summary, we have a sample of 50 (WISE-selected) and 22 (radio-selected), i.e., 72 radio sources that are selected without any dust bias and are suitable for 
the objectives of this survey.

\begin{longrotatetable}
\begin{deluxetable*}{ccccccccccccc}
	\tablecaption{Sample of radio sources for the uGMRT survey. }
\tabletypesize{\tiny}
\tablehead{
	\colhead{Source name} & \colhead{Flux} & \colhead{Obs. run} & \colhead{$W_1$} & \colhead{$W_2$} & \colhead{$W_3$} & \colhead{$z_{spec}$} & 
	\colhead{Beam} & \colhead{F$_p$} & \colhead{$\Delta$F} & \colhead{Beam} & \colhead{F$_p$} & \colhead{$\Delta$F} \\
	\vspace{-0.5cm} \\
	\colhead{           } & \colhead{(mJy)} & \colhead{             } & \colhead{ (mag) } & \colhead{ (mag) } & \colhead{ (mag) } & \colhead{   } & 
	\colhead{    } & \colhead{  (mJy\,b$^{-1}$)} & \colhead{ (mJy\,b$^{-1}$) } & \colhead{    } & \colhead{ (mJy\,b$^{-1}$)} & \colhead{(mJy\,b$^{-1}$)         } \\ 
	\vspace{-0.5cm} \\
	\colhead{(1) } & \colhead{ (2) } & \colhead{ (3)     } & \colhead{ (4)} & \colhead{ (5) } & \colhead{  (6)    } & \colhead{ (7)       } & \colhead{  (8)  } &
	\colhead{(9)} & \colhead{ (10)} & \colhead{ (11)} & \colhead{ (12)} & \colhead{(13)} \\
	\vspace{-0.7cm} \\
} 
\startdata
\vspace{-0.1cm} \\
\multicolumn{13}{c}{\scriptsize {WISE-selected targets}}   \\  
\vspace{-0.3cm} \\
J005408.40$-$033355.9 & 2210.8 &3,7      & 13.98  $\pm$ 0.03  & 13.08  $\pm$ 0.03  & 10.00  $\pm$ 0.07  & 0.210  &  $4.3^{\prime\prime}\times1.9^{\prime\prime},+56^\circ$  &  1674.0  & 2.3 & $6.1^{\prime\prime}\times3.3^{\prime\prime},+57^\circ$  & 2091.0  & 2.1\\   
J011815.36$-$012030.3 & 1248.9 &3,7      & 15.56  $\pm$ 0.05  & 14.28  $\pm$ 0.05  & 10.83  $\pm$ 0.13  & 1.162  &  $4.4^{\prime\prime}\times1.8^{\prime\prime},+58^\circ$  &  1024.0  & 2.0 & $6.0^{\prime\prime}\times3.2^{\prime\prime},+59^\circ$  & 1330.0  & 2.0 \\  
J011901.27$+$082954.7 & 2627.9 &3,7      & 14.19  $\pm$ 0.03  & 13.53  $\pm$ 0.03  & 10.68  $\pm$ 0.10  & 0.594  &  $4.7^{\prime\prime}\times1.9^{\prime\prime},+62^\circ$  &  2041.0  & 2.1 & $7.0^{\prime\prime}\times3.3^{\prime\prime},+70^\circ$  & 2526.0  & 3.0 \\  
J013212.20$-$065236.8 & 1258.9 &7        & 15.95  $\pm$ 0.05  & 14.73  $\pm$ 0.06  & 11.32  $\pm$ 0.16  & 1.511$^\dag$  &  $6.3^{\prime\prime}\times1.8^{\prime\prime},+62^\circ$  &  1026.0  & 2.1 &                             -                             &  -      &  -  \\  
J014012.82$-$095657.1 & 1175.3 &2,6      & 16.33  $\pm$ 0.07  & 15.31  $\pm$ 0.10  & 11.88  $\pm$ null  & -      &  $6.2^{\prime\prime}\times1.5^{\prime\prime},+57^\circ$  &  940.0   & 2.5 & $5.9^{\prime\prime}\times2.1^{\prime\prime},+55^\circ$  & 1141.0  & 2.6 \\ 
J070648.12$+$464756.8 & 1584.9 &2,3,6,7  & 16.46  $\pm$ 0.07  & 15.79  $\pm$ 0.14  & 11.77  $\pm$ 0.26  & -      &  $3.1^{\prime\prime}\times1.8^{\prime\prime},+58^\circ$  &  1536.0  & 2.1 & $3.9^{\prime\prime}\times2.0^{\prime\prime},+60^\circ$  & 1596.0  & 1.9 \\ 
J071338.15$+$434917.2 & 2011.4 &2,3,6,7  & 14.36  $\pm$ 0.03  & 13.63  $\pm$ 0.04  & 10.64  $\pm$ 0.10  & 0.518  &  $3.9^{\prime\prime}\times1.7^{\prime\prime},+64^\circ$  &  2007.0  & 2.4 & $3.9^{\prime\prime}\times2.2^{\prime\prime},+56^\circ$  & 2068.0  & 1.6 \\ 
J073555.54$+$330709.6 & 2473.1 &2,3,6,7  & 15.27  $\pm$ 0.04  & 14.54  $\pm$ 0.06  & 11.61  $\pm$ 0.25  & -      &  $2.9^{\prime\prime}\times1.8^{\prime\prime},+58^\circ$  &  2295.0  & 2.0 & $3.7^{\prime\prime}\times2.2^{\prime\prime},+57^\circ$  & 2274.0  & 1.8 \\ 
J082144.02$+$174820.5 & 1875.1 &2,3,6,7  & 14.16  $\pm$ 0.04  & 13.51  $\pm$ 0.03  &  9.80  $\pm$ 0.05  & 0.297  &  $3.0^{\prime\prime}\times1.7^{\prime\prime},+54^\circ$  &  1744.0  & 1.7 & $3.7^{\prime\prime}\times2.2^{\prime\prime},+54^\circ$  & 2029.0  & 1.9 \\ 
J082550.37$+$030924.8 & 1400.1 &2,3,6,7  & 11.90  $\pm$ 0.02  & 10.91  $\pm$ 0.02  &  7.59  $\pm$ 0.02  & 0.506  &  $3.3^{\prime\prime}\times1.9^{\prime\prime},+56^\circ$  &  1440.0  & 2.5 & $3.8^{\prime\prime}\times2.3^{\prime\prime},+53^\circ$  & 1593.0  & 2.1 \\ 
J085439.35$+$140552.1 & 2163.8 &1,2,3,4,7& 15.23  $\pm$ 0.04  & 14.10  $\pm$ 0.04  & 10.84  $\pm$ 0.12  & 1.02   &  $3.4^{\prime\prime}\times1.9^{\prime\prime},+55^\circ$  &  1302.0  & 2.4 & $3.8^{\prime\prime}\times2.3^{\prime\prime},+52^\circ$  & 1707.0  & 1.7 \\ 
J095828.78$-$013959.3 & 1213.7 &1,3,4,7  & 16.88  $\pm$ 0.11  & 15.87  $\pm$ 0.15  & 12.09  $\pm$ 0.44  & -      &  $5.0^{\prime\prime}\times2.1^{\prime\prime},-80^\circ$  &  932.0   & 2.0 & $4.2^{\prime\prime}\times3.0^{\prime\prime},+32^\circ$  & 1158.0  & 1.6 \\ 
J101145.46$+$462820.1 & 1557.2 &1,3,4,7  & 16.13  $\pm$ 0.06  & 15.39  $\pm$ 0.09  & 12.04  $\pm$ null  & 1.781  &  $6.1^{\prime\prime}\times1.7^{\prime\prime},-86^\circ$  &  1189.0  & 2.3 & $4.7^{\prime\prime}\times2.2^{\prime\prime},+78^\circ$  & 1151.0  & 2.1 \\ 
J104634.99$+$154347.2 & 1071.9 &1,3,4,7  & 16.63  $\pm$ 0.09  & 16.02  $\pm$ 0.20  & 12.55  $\pm$ null  & -      &  $5.3^{\prime\prime}\times1.9^{\prime\prime},-85^\circ$  &  949.0   & 2.2 & $4.4^{\prime\prime}\times2.4^{\prime\prime},-88^\circ$  & 1135.0  & 1.9 \\ 
J110808.31$+$143535.8 & 1348.7 &1,3,4,7  & 16.12  $\pm$ 0.06  & 15.20  $\pm$ 0.09  & 11.78  $\pm$ 0.29  & -      &  $3.3^{\prime\prime}\times1.8^{\prime\prime},+60^\circ$  &  1249.0  & 2.4 & $3.8^{\prime\prime}\times3.6^{\prime\prime},+53^\circ$  & 1433.0  & 1.7 \\ 
J110928.86$+$374431.4 & 1221.6 &1,4      & 16.60  $\pm$ 0.08  & 15.62  $\pm$ 0.12  & 11.66  $\pm$ 0.21  & 2.29   &  $4.0^{\prime\prime}\times1.7^{\prime\prime},+81^\circ$  &  997.0   & 1.8 & $3.9^{\prime\prime}\times2.4^{\prime\prime},+69^\circ$  & 893.0   & 1.9 \\ 
J110946.04$+$104343.4 & 1481.3 &1,4      & 15.65  $\pm$ 0.05  & 14.28  $\pm$ 0.05  & 10.71  $\pm$ 0.11  & -      &  $3.2^{\prime\prime}\times1.8^{\prime\prime},+78^\circ$  &  1383.0  & 2.0 & $2.7^{\prime\prime}\times2.1^{\prime\prime},+68^\circ$  & 1614.0  & 2.0 \\ 
J111917.36$-$052707.9 & 1174.4 &1,4      & 16.58  $\pm$ 0.09  & 15.94  $\pm$ 0.17  & 11.65  $\pm$ null  & 2.651$^\dag$  &  $3.9^{\prime\prime}\times2.0^{\prime\prime},+71^\circ$  &  884.0   & 2.1 & $2.6^{\prime\prime}\times2.3^{\prime\prime},+36^\circ$  & 945.0   & 2.0 \\  
J111925.22$-$030251.6 & 1730.4 &1,4      & 16.76  $\pm$ 0.10  & 15.96  $\pm$ 0.17  & 12.53  $\pm$ null  & 1.355  &  $3.6^{\prime\prime}\times1.8^{\prime\prime},+67^\circ$  &  1342.0  & 2.7 & $2.7^{\prime\prime}\times2.2^{\prime\prime},+37^\circ$  & 1498.0  & 2.2 \\ 
J113555.93$+$425844.8 & 1448.8 &1,4      & 16.32  $\pm$ 0.05  & 15.72  $\pm$ 0.09  & 12.61  $\pm$ 0.40  & -      &  $4.0^{\prime\prime}\times1.6^{\prime\prime},+78^\circ$  &  1390.0  & 1.8 & $3.6^{\prime\prime}\times2.1^{\prime\prime},+65^\circ$  & 1891.0  & 2.1 \\ 
J114049.54$+$591226.0 & 2179.4 &1,4      & 15.64  $\pm$ 0.04  & 14.55  $\pm$ 0.05  & 11.37  $\pm$ 0.14  & -      &  $3.4^{\prime\prime}\times2.0^{\prime\prime},+48^\circ$  &  1847.0  & 1.7 & $3.3^{\prime\prime}\times2.1^{\prime\prime},+15^\circ$  & 2571.0  & 2.2 \\ 
J115618.74$+$312805.0 & 2978.3 &1,4      & 13.74  $\pm$ 0.03  & 12.71  $\pm$ 0.02  &  9.48  $\pm$ 0.04  & 0.4171 &  $3.4^{\prime\prime}\times1.8^{\prime\prime},+62^\circ$  &  2812.0  & 2.3 & $2.8^{\prime\prime}\times2.2^{\prime\prime},+42^\circ$  & 3263.0  & 2.4 \\ 
J121514.69$+$173002.2 & 1010.2 &1,4      & 15.76  $\pm$ 0.05  & 15.04  $\pm$ 0.08  & 12.00  $\pm$ 0.28  & -      &  $3.1^{\prime\prime}\times1.8^{\prime\prime},+62^\circ$  &  901.0   & 2.1 & $3.7^{\prime\prime}\times2.1^{\prime\prime},+57^\circ$  & 1011.0  & 2.4 \\ 
J122758.78$+$363511.6 & 2098.4 &1,4      & 17.50  $\pm$ 0.16  & 16.82  $\pm$ null  & 12.05  $\pm$ null  & 1.975  &  $3.1^{\prime\prime}\times1.9^{\prime\prime},+59^\circ$  &  2021.0  & 2.1 & $3.6^{\prime\prime}\times2.3^{\prime\prime},+51^\circ$  & 2154.0  & 2.1 \\ 
J134733.42$+$121724.1$^\ddag$ & 5397.2 &1,4      & 11.52  $\pm$ 0.03  & 10.23  $\pm$ 0.03  &  6.29  $\pm$ 0.03  & 0.121  &  $3.3^{\prime\prime}\times1.9^{\prime\prime},+66^\circ$  &  5082.0  & 2.9 & $3.7^{\prime\prime}\times2.1^{\prime\prime},+58^\circ$  & 5599.0  & 2.8 \\  
J141043.03$+$364721.9 & 1226.1 &1,5      & 16.81  $\pm$ 0.08  & 15.72  $\pm$ 0.11  & 12.27  $\pm$ 0.26  &  -     &  $3.9^{\prime\prime}\times1.9^{\prime\prime},+55^\circ$  &  1531.0  & 2.6 & $3.7^{\prime\prime}\times2.4^{\prime\prime},+52^\circ$  & 1515.0  & 2.3 \\ 
J141604.18$+$344436.5 & 1863.7 &1,5      & 17.28  $\pm$ 0.12  & 16.02  $\pm$ 0.14  & 12.40  $\pm$ 0.30  &  -     &  $3.9^{\prime\prime}\times1.9^{\prime\prime},+55^\circ$  &  2235.0  & 2.9 & $3.7^{\prime\prime}\times2.3^{\prime\prime},+54^\circ$  & 2106.0  & 2.2 \\ 
J142105.73$+$414449.7 & 3146.9 &1,5      & 15.29  $\pm$ 0.04  & 14.52  $\pm$ 0.05  & 10.77  $\pm$ 0.07  &  0.367 &  $4.0^{\prime\prime}\times2.1^{\prime\prime},+50^\circ$  &  2990.0  & 2.8 & $3.7^{\prime\prime}\times2.6^{\prime\prime},+49^\circ$  & 3017.0  & 2.9 \\ 
J142921.93$+$540611.3 & 1028.3 &1,5      & 16.32  $\pm$ 0.05  & 15.26  $\pm$ 0.07  & 12.04  $\pm$ 0.21  &  3.03  &  $3.8^{\prime\prime}\times2.5^{\prime\prime},+35^\circ$  &  926.0   & 2.2 & $3.5^{\prime\prime}\times3.2^{\prime\prime},+17^\circ$  & 1119.0  & 2.2 \\ 
J144343.22$+$503431.7 & 1209.6 &1,5      & 15.49  $\pm$ 0.04  & 14.32  $\pm$ 0.04  & 11.04  $\pm$ 0.10  &  -     &  $3.9^{\prime\prime}\times2.5^{\prime\prime},+41^\circ$  &  1025.0  & 2.2 & $3.5^{\prime\prime}\times3.2^{\prime\prime},+54^\circ$  & 1257.0  & 2.2 \\ 
J151340.20$+$233835.3 & 1767.5 &1,5      & 16.13  $\pm$ 0.05  & 14.77  $\pm$ 0.05  & 11.27  $\pm$ 0.12  &  -     &  $4.8^{\prime\prime}\times2.0^{\prime\prime},+57^\circ$  &  1698.0  & 2.2 & $4.5^{\prime\prime}\times2.4^{\prime\prime},+60^\circ$  & 1832.0  & 2.1 \\ 
J151656.61$+$183021.6 & 1335.2 &1,5      & 15.53  $\pm$ 0.04  & 14.26  $\pm$ 0.04  & 10.24  $\pm$ 0.05  & 5.5    &  $5.3^{\prime\prime}\times1.8^{\prime\prime},+59^\circ$  &  1191.0  & 2.8 & $5.1^{\prime\prime}\times2.6^{\prime\prime},+60^\circ$  & 1266.0  & 2.4 \\ 
J152114.51$+$043020.0 & 3927.2 &1,5      & 15.86  $\pm$ 0.05  & 14.67  $\pm$ 0.06  & 11.11  $\pm$ 0.10  & 1.294  &  $7.7^{\prime\prime}\times1.7^{\prime\prime},+61^\circ$  &  3238.0  & 2.9 & $7.2^{\prime\prime}\times2.5^{\prime\prime},+61^\circ$  & 3875.0  & 2.5 \\ 
J160207.27$+$332653.1 & 2990.6 &1,5      & 15.72  $\pm$ 0.04  & 14.43  $\pm$ 0.04  & 10.93  $\pm$ 0.07  & 1.1    &  $5.0^{\prime\prime}\times2.3^{\prime\prime},+56^\circ$  &  2275.0  & 2.3 & $4.8^{\prime\prime}\times3.0^{\prime\prime},+64^\circ$  & 2463.0  & 2.4 \\ 
J160846.13$+$102908.2 & 1392.0 &1,5      & 13.17  $\pm$ 0.14  & 12.10  $\pm$ 0.02  &  8.63  $\pm$ 0.03  & 1.232  &  $8.4^{\prime\prime}\times1.8^{\prime\prime},+62^\circ$  &  746.0   & 2.2 & $7.9^{\prime\prime}\times2.4^{\prime\prime},+66^\circ$  & 894.0   & 2.1 \\ 
J161219.02$+$222215.6 & 1401.9 &5        & 15.16  $\pm$ 0.04  & 14.30  $\pm$ 0.04  & 10.82  $\pm$ 0.10  & 0.6362 &  $9.6^{\prime\prime}\times2.1^{\prime\prime},+63^\circ$  &  742.0   & 5.1 &                   -                                          &  -      &  -  \\   
J162557.66$+$413441.2 & 1677.4 &1,5      & 16.40  $\pm$ 0.06  & 15.27  $\pm$ 0.07  & 12.00  $\pm$ 0.24  & 2.55   &  $6.5^{\prime\prime}\times2.8^{\prime\prime},+60^\circ$  &  1189.0  & 3.2 & $6.8^{\prime\prime}\times3.0^{\prime\prime},+73^\circ$  & 1436.0  & 2.4 \\ 
J163433.86$+$624535.7 & 5001.9 &1,5      & 15.82  $\pm$ 0.04  & 14.99  $\pm$ 0.05  & 10.68  $\pm$ 0.05  & 0.988  &  $5.5^{\prime\prime}\times3.7^{\prime\prime},+87^\circ$  &  4120.0  & 2.8 & $6.7^{\prime\prime}\times3.2^{\prime\prime},+88^\circ$  & 5292.0  & 3.0 \\ 
J163822.12$+$103507.9 & 1134.1 &3,7      & 16.60  $\pm$ 0.08  & 15.60  $\pm$ 0.11  & 11.93  $\pm$ null  & -      &  $3.6^{\prime\prime}\times1.9^{\prime\prime},+83^\circ$  &  1008.0  & 2.0 & $4.0^{\prime\prime}\times3.2^{\prime\prime},+60^\circ$  & 1055.0  & 1.7 \\ 
J164047.99$+$122002.2 & 2070.1 &3,7      & 16.08  $\pm$ 0.06  & 14.76  $\pm$ 0.06  & 11.23  $\pm$ 0.16  & 1.152  &  $3.3^{\prime\prime}\times1.8^{\prime\prime},+78^\circ$  &  2113.0  & 2.2 & $3.8^{\prime\prime}\times3.0^{\prime\prime},+52^\circ$  & 1825.0  & 1.8 \\ 
J205828.84$+$054250.7 & 1213.2 &1,2,5,6  & 17.76  $\pm$ 0.21  & 16.81  $\pm$ 0.31  & 12.65  $\pm$ null  & 1.381  &  $5.0^{\prime\prime}\times1.6^{\prime\prime},+58^\circ$  &  1068.0  & 3.0 & $5.8^{\prime\prime}\times2.2^{\prime\prime},+56^\circ$  & 1165.0  & 2.1 \\ 
J212339.12$-$011234.3 & 1086.6 &1,2,5,6  & 16.98  $\pm$ 0.11  & 16.34  $\pm$ 0.23  & 12.41  $\pm$ null  & 1.158  &  $4.9^{\prime\prime}\times1.8^{\prime\prime},+58^\circ$  &  870.0   & 2.1 & $6.0^{\prime\prime}\times2.0^{\prime\prime},+59^\circ$  & 1053.0  & 2.2 \\ 
J213032.84$+$050217.5 & 4098.0 &1,2,5,6  & 16.20  $\pm$ 0.06  & 15.51  $\pm$ 0.13  & 12.16  $\pm$ null  & 0.99   &  $5.1^{\prime\prime}\times1.7^{\prime\prime},+58^\circ$  &  3576.0  & 3.0 & $4.8^{\prime\prime}\times2.1^{\prime\prime},+56^\circ$  & 3834.0  & 2.4 \\ 
J213638.57$+$004154.5 & 3472.5 &2,6      & 13.65  $\pm$ 0.03  & 12.65  $\pm$ 0.03  &  9.41  $\pm$ 0.05  & 1.94   &  $9.1^{\prime\prime}\times1.7^{\prime\prime},+62^\circ$  &  2478.0  & 3.2 & $8.6^{\prime\prime}\times2.2^{\prime\prime},+63^\circ$  & 2264.0  & 2.8 \\ 
J215023.59$+$144947.5 & 2666.7 &1,2,5,6  & 15.60  $\pm$ 0.05  & 14.81  $\pm$ 0.06  & 11.31  $\pm$ 0.15  & 0.4    &  $4.5^{\prime\prime}\times1.8^{\prime\prime},+58^\circ$  &  2334.0  & 2.2 & $5.6^{\prime\prime}\times2.2^{\prime\prime},+55^\circ$  & 2435.0  & 2.4 \\ 
J220620.82$+$055611.8 & 1064.7 &2,6      & 16.73  $\pm$ 0.09  & 15.78  $\pm$ 0.14  & 12.49  $\pm$ 0.53  & -      &  $8.6^{\prime\prime}\times1.6^{\prime\prime},+63^\circ$  &  774.0   & 2.8 & $8.1^{\prime\prime}\times2.2^{\prime\prime},+64^\circ$  & 1024.0  & 2.5 \\ 
J222547.28$-$045701.9$^*$ & 7410.6 &-        & 12.82  $\pm$ 0.02  & 11.63  $\pm$ 0.02  &  8.15  $\pm$ 0.02  & 1.404  &               -                                          &    -     & -   &              -                                               &  -      & -   \\   
J232236.09$+$081201.3 & 1184.6 &2        & 15.08  $\pm$ 0.04  & 14.07  $\pm$ 0.05  & 10.63  $\pm$ 0.10  & 2.09   &               -                                          &    -     & -   & $6.1^{\prime\prime}\times2.3^{\prime\prime},+62^\circ$  & 1066.0  & -   \\  
J232510.23$-$034446.7 & 1223.5 &2,6      & 16.37  $\pm$ 0.08  & 15.17  $\pm$ 0.10  & 11.54  $\pm$ 0.21  & 1.509  &  $8.9^{\prime\prime}\times1.6^{\prime\prime},+62^\circ$  &  842.0   & 2.3 & $8.6^{\prime\prime}\times2.1^{\prime\prime},+63^\circ$  &  948.0  & 2.1 \\    
J234058.05$+$043114.9 & 1612.9 &2,6      & 14.84  $\pm$ 0.03  & 13.95  $\pm$ 0.05  & 10.49  $\pm$ 0.09  & 2.589  &  $8.7^{\prime\prime}\times1.7^{\prime\prime},+62^\circ$  &  1377.0  & 2.4 & $8.4^{\prime\prime}\times2.3^{\prime\prime},+65^\circ$  & 1376.0  & 3.5 \\    
\vspace{-0.1cm} \\
\multicolumn{13}{c}{\scriptsize {Radio-selected targets (gain calibrators)}}   \\                     
\vspace{-0.3cm} \\
J005905.53$+$000651.5 &  2508.8 & 2,3,6,7 & 13.31  $\pm$ 0.03  & 12.14  $\pm$ 0.02  &  9.33  $\pm$ 0.04  & 0.719 &  $4.6^{\prime\prime}\times1.9^{\prime\prime},+59^\circ$ & 2153.0 & 3.5  &  $5.9^{\prime\prime}\times2.2^{\prime\prime},+58^\circ$  & 2647.0  & 6.9 \\  
J074110.70$+$311200.4 &  2284.3 & 2,3,6,7 & 12.35  $\pm$ 0.02  & 11.26  $\pm$ 0.02  &  8.69  $\pm$ 0.03  & 0.632 &  $3.0^{\prime\prime}\times1.8^{\prime\prime},+60^\circ$ & 2058.0 & 5.2  &  $4.1^{\prime\prime}\times2.9^{\prime\prime},+35^\circ$  & 2068.0  & 4.6 \\  
J080518.22$+$614423.2 &  828.2  & 2,3,6,7 & 15.79  $\pm$ 0.05  & 14.58  $\pm$ 0.05  & 10.99  $\pm$ 0.12  & 3.033 &  $3.2^{\prime\prime}\times2.0^{\prime\prime},+34^\circ$ & 874.0  & 1.7  &  $4.9^{\prime\prime}\times2.9^{\prime\prime},-6^\circ$   &  942.0  & 1.7 \\   
J083454.91$+$553421.0 &  8283.1 & 2,3,6,7 & 13.72  $\pm$ 0.03  & 13.43  $\pm$ 0.03  & 10.52  $\pm$ 0.08  & 0.268 &  $3.3^{\prime\prime}\times2.0^{\prime\prime},+50^\circ$ & 8253.0 & 5.8  &  $4.7^{\prime\prime}\times2.9^{\prime\prime},+21^\circ$  &  8320.0 & 7.5 \\  
J084205.09$+$183541.8 &  1259.7 & 1-4,6,7 & 13.46  $\pm$ 0.03  & 12.10  $\pm$ 0.02  &  9.20  $\pm$ 0.05  & 1.270 &  $3.1^{\prime\prime}\times1.9^{\prime\prime},+58^\circ$ & 1102.0 & 4.3  &  $3.7^{\prime\prime}\times2.2^{\prime\prime},+57^\circ$  &  1163.0 & 2.8 \\  
J094336.86$-$081932.0 &  2756.2 & 1,3,4,7 & 14.45  $\pm$ 0.03  & 14.06  $\pm$ 0.04  & 11.62  $\pm$ 0.24  & 0.228 &  $5.5^{\prime\prime}\times2.1^{\prime\prime},-77^\circ$ & 2163.0 & 5.7  &  $7.0^{\prime\prime}\times2.5^{\prime\prime},-70^\circ$   &  2790.0 & 5.7 \\ 
J103507.04$+$562847.3 &  1801.9 & 1,3,4,7 & 15.35  $\pm$ 0.04  & 14.35  $\pm$ 0.04  & 11.43  $\pm$ 0.19  & 0.460 &  $3.4^{\prime\prime}\times2.1^{\prime\prime},+86^\circ$ & 1584.0 & 3.0  &  $4.4^{\prime\prime}\times3.0^{\prime\prime},-5^\circ$   &  1824.0 & 4.6 \\  
J111120.09$+$195536.1 &  1194.8 & 1,3,4,7 & 14.89  $\pm$ 0.03  & 14.73  $\pm$ 0.07  & 11.89  $\pm$ null  & 0.299 &  $7.2^{\prime\prime}\times1.7^{\prime\prime},-79^\circ$ & 1047.0 & 5.2  &  $5.9^{\prime\prime}\times2.3^{\prime\prime},-83^\circ$   &  1260.0 & 5.3 \\ 
J112027.81$+$142054.4 &  2446.9 & 1,4     & 15.89  $\pm$ 0.05  & 15.48  $\pm$ 0.12  & 11.88  $\pm$ null  & 0.363 &  $3.6^{\prime\prime}\times1.8^{\prime\prime},+81^\circ$ & 2395.0 & 4.9  &  $3.2^{\prime\prime}\times2.4^{\prime\prime},+73^\circ$  &  2511.0 & 4.9 \\  
J114658.31$+$395834.9 &   330.9 & 1,3,4,7 & 12.48  $\pm$ 0.02  & 11.38  $\pm$ 0.02  &  8.55  $\pm$ 0.03  & 1.091 &  $4.8^{\prime\prime}\times1.7^{\prime\prime},+88^\circ$ & 858.0  & 5.3  &  $4.3^{\prime\prime}\times2.2^{\prime\prime},+78 ^\circ$  &   870.0 & 5.3 \\ 
J115043.88$-$002354.3 &  2773.9 & 1,4     & 14.01  $\pm$ 0.03  & 13.13  $\pm$ 0.03  & 10.09  $\pm$ 0.07  & 1.978 &  $4.5^{\prime\prime}\times2.1^{\prime\prime},+79^\circ$ & 2629.0 & 4.6  &  $3.0^{\prime\prime}\times2.6^{\prime\prime},+76^\circ$  &  2397.0 & 5.3 \\  
J122758.78$+$363511.6 &  2098.4 & 1,4     & 17.50  $\pm$ 0.16  & 16.82  $\pm$ null  & 12.05  $\pm$ null  & 1.975 &  $3.5^{\prime\prime}\times1.8^{\prime\prime},+66^\circ$ & 2087.0 & 5.1  &  $2.9^{\prime\prime}\times2.2^{\prime\prime},+47^\circ$  &  2102.0 & 5.4 \\  
J125438.28$+$114106.4 &   792.8 & 1,4     & 15.15  $\pm$ 0.04  & 14.92  $\pm$ 0.07  & 12.21  $\pm$ null  &   -   &  $3.4^{\prime\prime}\times1.9^{\prime\prime},+68^\circ$ & 671.0  & 5.9  &  $3.8^{\prime\prime}\times2.2^{\prime\prime},+61^\circ$  &   613.0 & 6.0 \\  
J140501.10$+$041536.2 &  933.0 &  1,4     & 15.72  $\pm$ 0.05  & 14.64  $\pm$ 0.05  & 11.46  $\pm$ 0.14  & 3.215 &  $3.3^{\prime\prime}\times1.9^{\prime\prime},+64^\circ$ & 857.0  & 2.1  &  $3.9^{\prime\prime}\times2.2^{\prime\prime},+57^\circ$  &   976.0 & 1.9 \\   
J143844.71$+$621154.5 &  2410.4 & 1,5     & 14.72  $\pm$ 0.03  & 13.31  $\pm$ 0.03  & 10.35  $\pm$ 0.05  & 1.094 &  $3.9^{\prime\prime}\times2.6^{\prime\prime},+26^\circ$ & 2377.0 & 6.3  &  $3.8^{\prime\prime}\times3.1^{\prime\prime},-5^\circ$   &   2875.0& 5.8 \\  
J144516.48$+$095836.0 &  2417.6 & 1,5     & 14.84  $\pm$ 0.03  & 14.26  $\pm$ 0.04  & 10.65  $\pm$ 0.07  & 3.541 &  $8.3^{\prime\prime}\times1.9^{\prime\prime},+61^\circ$ & 1887.0 & 6.1  &  $7.9^{\prime\prime}\times2.6^{\prime\prime},+63^\circ$  &   2179.0& 5.9 \\  
J150609.50$+$373051.4 &   937.1 & 1,5     & 13.90  $\pm$ 0.03  & 12.55  $\pm$ 0.02  &  9.65  $\pm$ 0.03  & 0.673 &  $4.1^{\prime\prime}\times2.0^{\prime\prime},+56^\circ$ & 1062.0 & 6.5  &  $3.8^{\prime\prime}\times2.4^{\prime\prime},+51^\circ$  &    978.0& 5.4 \\  
J160913.31$+$264129.2 &  4908.2 & 5       & 15.30  $\pm$ 0.04  & 15.02  $\pm$ 0.06  & 12.34  $\pm$ null  & 0.474 &  $8.5^{\prime\prime}\times2.3^{\prime\prime},+61^\circ$ & 3550.0 & 5.7  &  -                                                          &     -   &  -  \\ 
J163515.51$+$380804.8 &  2726.0 & 1,5     & 12.68  $\pm$ 0.07  & 11.51  $\pm$ 0.02  &  8.54  $\pm$ 0.02  & 1.814 &  $6.2^{\prime\prime}\times2.7^{\prime\prime},+58^\circ$ & 1986.0 & 6.3  &  $6.0^{\prime\prime}\times3.0^{\prime\prime},+70^\circ$  &  2419.0 & 5.7 \\    
J164047.99$+$122002.2 &  2070.1 & 1,5     & 16.08  $\pm$ 0.06  & 14.76  $\pm$ 0.06  & 11.23  $\pm$ 0.16  & 1.152 &  $6.2^{\prime\prime}\times1.9^{\prime\prime},+59^\circ$ & 1920.0 & 9.6  &  $5.8^{\prime\prime}\times2.4^{\prime\prime},+61^\circ$  &  2322.0 & 6.5 \\    
J221237.97$+$015251.7 &  2809.8 & 2,6     & 15.27  $\pm$ 0.04  & 13.97  $\pm$ 0.04  & 10.82  $\pm$ 0.12  & 1.126 &  $7.9^{\prime\prime}\times1.7^{\prime\prime},+61^\circ$ & 2370.0 & 7.6  &  $7.7^{\prime\prime}\times2.1^{\prime\prime},+62^\circ$  &  3150.0 & 7.6 \\    
J233040.83$+$110018.3 &  1203.8 & 2,6     & 14.85  $\pm$ 0.03  & 13.52  $\pm$ 0.04  & 10.38  $\pm$ 0.08  & 1.502 &  $6.2^{\prime\prime}\times1.8^{\prime\prime},+59^\circ$ & 1077.0 & 7.8  &  $5.9^{\prime\prime}\times2.4^{\prime\prime},+62^\circ$  &  1139.0 & 5.9  \\   
\label{tab:wisesamp}
\enddata
\tablecomments{ 
\scriptsize{ Column 1: NVSS ID based on Right Ascension and Declination (J2000). Column 2: 20\,cm flux density from NVSS. Column 3: observing run(s) - see Table~\ref{tab:obslog}. 
Columns 4-6: WISE $W_1$, $W_2$ and $W_3$ magnitudes (mag) with errors. Column 7: spectroscopic redshift from literature.  Columns 8 and 9: Synthesized beam and peak flux density (mJy\,beam$^{-1}$) for the images from spectral window covering 1310-1370MHz.  Column 10: spectral rms (mJy\,beam$^{-1}$) at 1355\,MHz. 
Column 11-12: same as 8 and 9, but for the spectral window covering 1050-1110\,MHz.  Column 13: spectral rms (mJy\,beam$^{-1}$)  at 1075\,MHz.\\
$^\dag$: Redshift from MALS-SALT/NOT survey \citep[][]{Krogager18}.
$^\ddag$: Known associated \hi\ 21-cm absorber, but the relevant frequency range excluded due to strong RFI \citep[][]{Mirabel89}.
$^*$: Could not be scheduled for observations.
}
}
\end{deluxetable*}
\end{longrotatetable}


\subsection{uGMRT Observations and data analysis}      
\label{sec:obs}   
%
\begin{figure*} 
\includegraphics[trim = {0cm 0cm 0cm 0cm}, width=1.0\textwidth,angle=0]{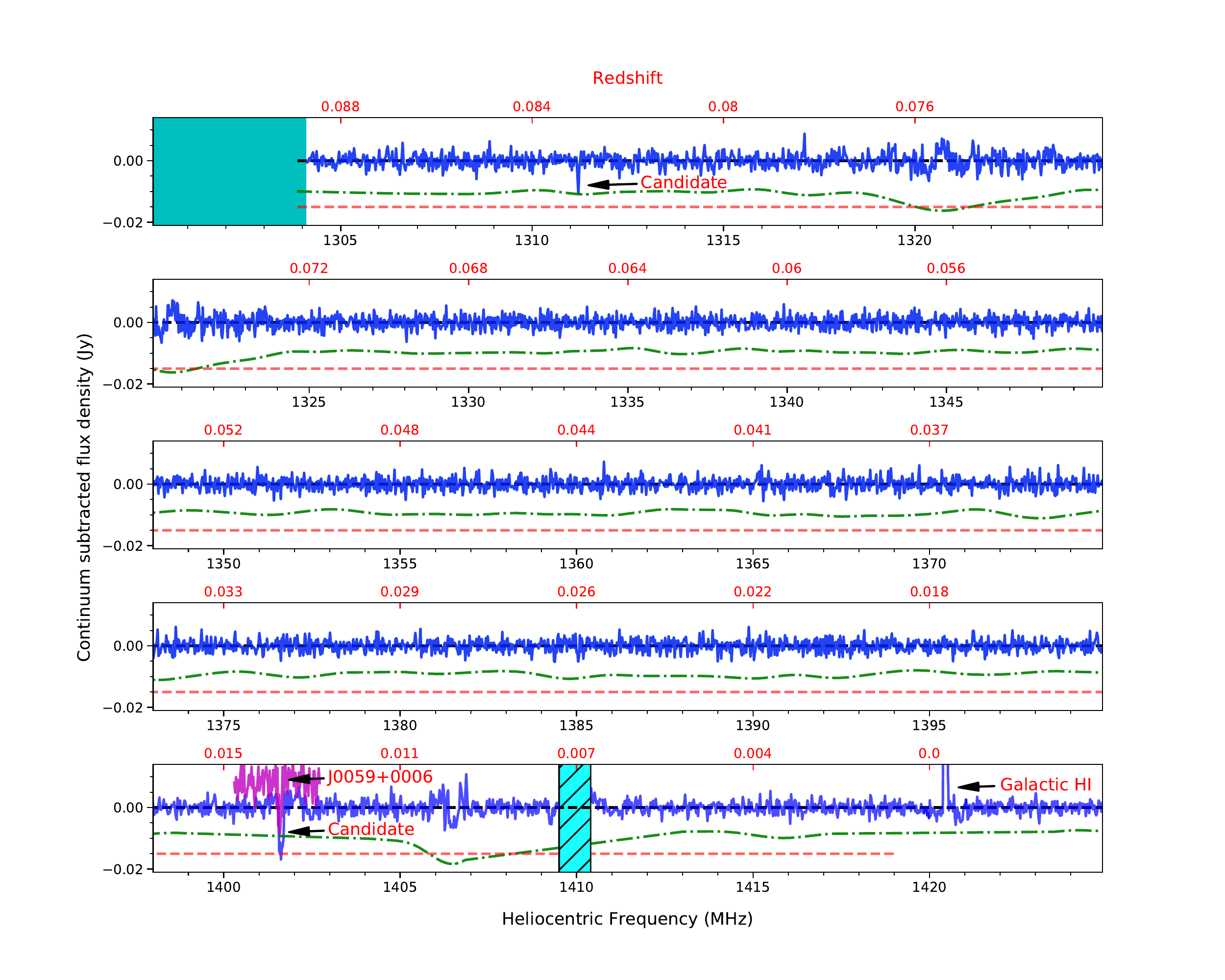}  
\vskip+0.0cm  
\caption{
	The continuum subtracted unsmoothed Stokes-$I$ spectrum of J0118-0120.  The {\it x}-axis is (bottom) heliocentric frequency and (top) the corresponding 
	\hi\ 21-cm line redshift. 
	Shaded regions mark frequency ranges that were masked 
	prior to any calibration.  Hatched regions were masked after calibration and imaging to exclude invalid spectral line features identified 
	during the absorption line analysis.  The dash-dotted line is 5$\times\sigma_{\rm rolling}$.
	 The dashed line marks the frequency range contributing to the redshift path. Note that the absorption feature at 1401.5\,MHz is also present in 
	 the spectra of J0059$+$0006 and, hence, marked spurious (see text for details).
} 
\label{fig:lbhigh}   
\end{figure*} 
%
\begin{deluxetable}{cccc}
\tabletypesize{\small}
\tablecaption{Details of uGMRT observations.}
\tablehead{
\colhead{ Obs. run} &  \colhead{Freq. setup      } & \colhead{Date} & \colhead{Duration} \\
			 &  \colhead{        (MHz)    } &                & \colhead{(hours)} 
}
\startdata
   1           &  1000-1200    & 2018 January 12       &  18.5    \\
   2           &      ''       & 2018 January 15       &  10.5    \\
   3           &      ''       & 2018 February 01      &  11.3    \\
   4           &  1260-1460    & 2018 January 11       &  10.8    \\
   5           &      ''       & 2018 January 12       &  10.0    \\
   6           &      ''       & 2018 January 14       &  10.7    \\
   7           &      ''       & 2018 January 30       &  11.8    \\
\label{tab:obslog}
\enddata
\end{deluxetable}
%
%
We used {\tt Band-5} (1000-1450\,MHz) of the uGMRT to observe the sample. The total allocated time of $\sim$85\,hours was spread over 
seven observing runs in January - February 2018.  The observations are summarized in Table~\ref{tab:obslog}. 
For all the observations, the uGMRT Wideband Backend (GWB) was used to configure a baseband bandwidth of 200\,MHz split into 8192 frequency channels.  This configuration yielded a 
channel resolution of 24.414\,kHz, which at 1200\,MHz corresponds to a velocity resolution of 6\,\kms.  
Runs 1-3 covered the frequency range of 1000-1200\,MHz, which for \hi\ corresponds to a redshift range of $0.18<z<0.42$. 
The corresponding coverage for the OH main line at 1667.359\,MHz is $0.39<z<0.67$.
Runs 4-7 covered 1260-1460\,MHz, which corresponds to $0<z<0.13$ ($0.14<z<0.32$) for the 21-cm line (OH main line).

\begin{figure*} 
\centerline{\vbox{
\centerline{\hbox{ 
\includegraphics[trim = {0cm 0cm 0cm 0cm}, width=1.0\textwidth,angle=0]{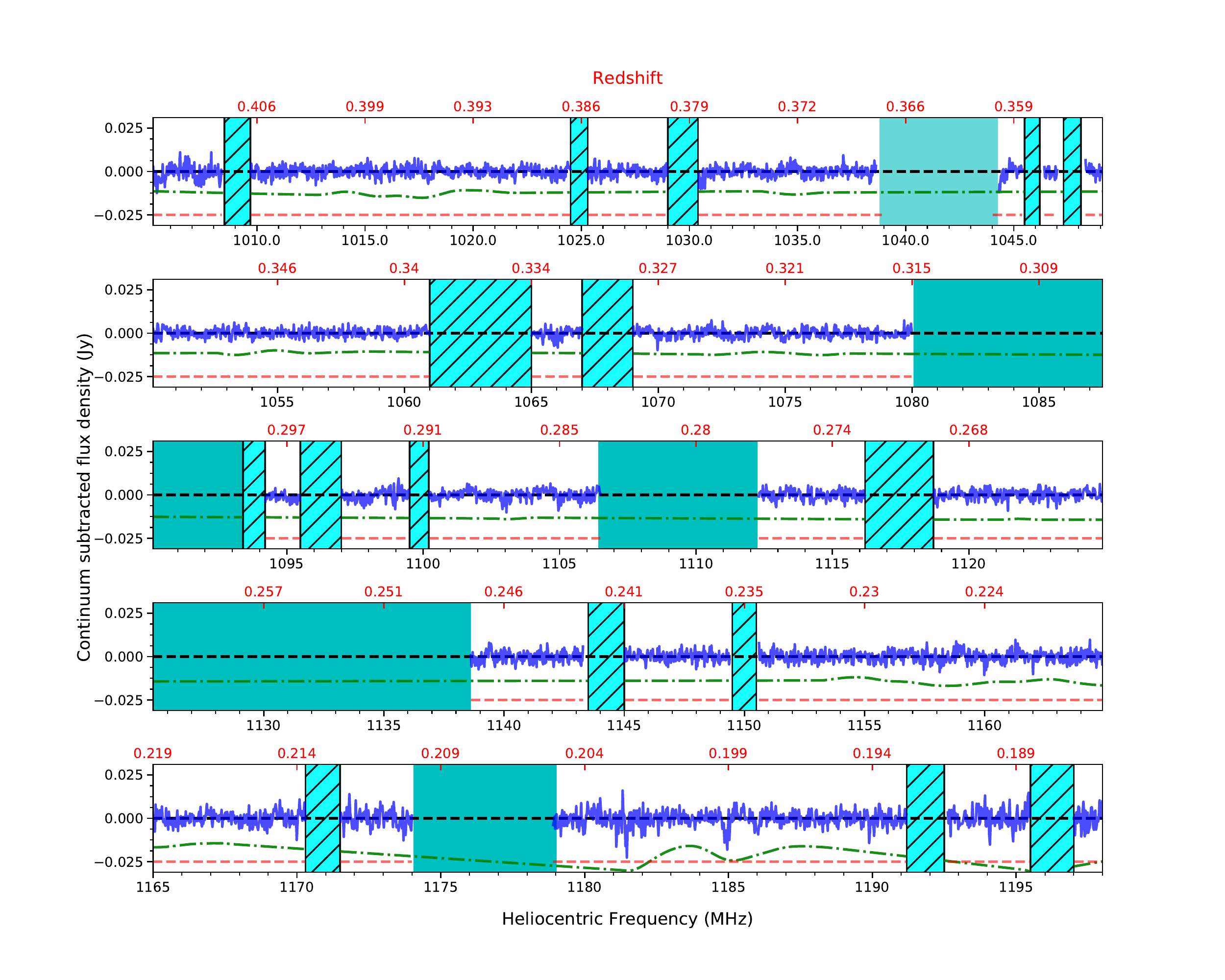}  
}} 
}}  
\vskip+0.0cm  
\caption{
	The continuum subtracted unsmoothed Stokes-$I$ spectrum of J1140+5912 (see Fig.~\ref{fig:lbhigh} caption for details). 
	The feature at 1185.0\,MHz is detected as a candidate in the smoothed spectra but fails other 
	tests to be classified as a true absorption.
} 
\label{fig:lblow}   
\end{figure*} 
%
During each observing run 3C48, 3C147, and/or 3C286 were observed for flux density and bandpass calibrations.  
If required, a complex gain calibrator was also observed  for a target source.
To minimize the calibration overheads, the observations were optimized to use the same gain 
calibrator for as many target sources as possible. 
As noted in the previous section, the observations of gain calibrators led to an independent radio-selected sample of 22 sources that were also searched  
for intervening absorption.  

In a given observing run, a target source was visited only once for a duration of $\sim$30\,minutes. The details of each run are summarized in Table 2, 
and the details of runs associated with each target source are listed in column\,3 of Table~\ref{tab:wisesamp}.  
It may be noted that some of the target sources were observed during more than one run.  
In general, the on-source time for gain calibrators per observing run is 5-10\,minutes (also see column\,3 of Table~\ref{tab:wisesamp}). 
%

The data were edited, calibrated, and imaged using {\tt ARTIP} following the steps described in Section~\ref{sec:artip}.  The {\tt ARTIP-CAL} package was 
used to generate {\tt corrected} visibilities for all the sources in the sample.  
The initial RFI mask applied prior to any calibration is shown as shaded regions in Figs.~\ref{fig:lbhigh} and \ref{fig:lblow}.

The spectral line processing through {\tt ARTIP-CUBE} was streamlined and sped up by partitioning the 200\,MHz bandwidth into four spectral 
windows with an overlap of 10\,MHz between adjacent windows. 
For the 1000-1200\,MHz setup, these spectral windows covered the following frequency ranges: 1000-1060, 1050-1110, 1100-1160, and 1150-1200\,MHz.
For the 1260-1460\,MHz setup, they covered: 1260-1320, 1310-1370, 1360-1420, and 1410-1460\,MHz.
The {\tt corrected} visibilities (measurement set) for each spectral window were processed separately (independently) for RFI flagging, continuum imaging with  
self-calibration, and continuum subtraction.
No broadband continuum image was generated (see Section~\ref{sec:pkscont}).
The synthesized beams and continuum peak flux densities for the images generated for 1310-1370\,MHz and 1050-1110\,MHz are provided in 
columns 8-9 and 11-12 of Table~\ref{tab:wisesamp}, respectively.

The spectral line processing through stage (4) of {\tt ARTIP-CUBE} proceeded exactly the same way as in Section~\ref{sec:pkscube}.
Only observing runs 1 and 4 were processed through {\tt ARTIP} in an interactive way.  The optimized configurations obtained from these were used to process 
data from all the remaining runs in a completely non-interactive manner. As noted previously, {\tt ARTIP} is fully capable of managing 
arbitrary combinations of calibrators and target sources present in a measurement set.

Finally, the target source XX and YY spectra in the heliocentric frame were extracted from image cubes and corrected for residual errors due to continuum 
subtraction and bandpass calibration. 
The two parallel hand spectra and channels in the overlapping frequency regions  were then combined to form the final Stokes-$I$ spectrum covering the full band. 
As examples, we present unsmoothed spectra for J0118-0120 (Run 4) and J1140+5912 (Run 1) in Figs.~\ref{fig:lbhigh} and \ref{fig:lblow}, respectively. 
Since we reach adequate column density sensitivity from individual runs, we do not combine spectra from multiple runs. Instead, as described below, 
multiple spectra are used for reliable identification of true spectral line features.
The $\sigma_{\rm rolling}$ for the best Stokes-$I$ spectra at 1355\,MHz and 1075\,MHz are provided in columns\,10 and 13 of 
Table~\ref{tab:wisesamp}, respectively.

We searched the Stokes-$I$ spectra for absorption lines adopting a slightly modified approach compared to Section~\ref{sec:pks}. The algorithm requires at a channel $j$  
(i) flux density F($j$) $<$$-5\times\sigma_{rolling}$($j$),  
and (ii) heliocentric frequency $\nu$($j$) $\ge \nu_o$/(1 + $z_{spec}$), where $\nu_o$ is the rest-frame \hi\ 21-cm or OH 18-cm line frequency. 
For an absorption candidate to be deemed real, we applied two additional conditions: the absorption is (i) reproduced in spectra from multiple runs, and (ii) not present in 
any spectra of other sources from the same observing run. For example, in the case of J0118-0120, the algorithm identified 
three statistically 
significant candidates at 1311.2, 1401.5, and 1410.1\,MHz in the unsmoothed spectrum (resolution$\sim$6\,\kms\ at 1200\,MHz) presented in Fig.~\ref{fig:lbhigh}.  
All three are due to sporadic RFI.  
To illustrate this pattern for the feature at 1401.5\,MHz, in Fig.~\ref{fig:lbhigh} we also plot the spectra of J0059$+$0006, the gain calibrator used for J0118-0120.
The hatched regions in Figs.~\ref{fig:lbhigh} and \ref{fig:lblow} mask frequency ranges with invalid spectral line features identified through the 
detection algorithm. The features at 1311.2 and 1401.5 are not hatched to illustrate the identification process presented above. 

We also smoothed all the spectra by 4, 8, and 16 channels (i.e. resolutions of 24, 48, and 96\,\kms, respectively) and searched for absorption candidates.
Overall, we do not detect any valid absorption candidates in the unsmoothed and smoothed spectra for any of the radio sources in the sample.

\section{Results}    
\label{sec:res}  

In this section we first compute the upper limits on the incidences, i.e., the numbers per unit redshift of intervening \hi\ 21-cm and OH 18-cm lines from 
the uGMRT survey.  Then we compare these with the expectations from \mgii\ absorption-based 
21-cm line searches and the properties of luminous galaxies within 150\,kpc of the sight lines towards the radio sources. 
In several cases, the redshifted \hi\ and OH line frequencies associated with the radio sources lie within the 
uGMRT band, allowing a search for associated absorption whose results are
summarized in the last subsection.

\subsection{Incidence of intervening \hi\ and OH absorption}    
\label{sec:dndz}  

The starting point of computing the incidence of absorption lines is to determine the sensitivity function, 
$g({\cal{T}}, z)$, as a function of integrated optical depth (${\cal{T}}$) and redshift ($z$). 
This function represents the number of spectra in which it is possible to detect absorption of a given strength at a given redshift.

Adapting the well-established formalism for estimating the completeness of absorption line surveys 
\citep[see, for example,][]{Lanzetta87}, we define the sensitivity function at the $j$th ${\cal{T}}$-interval and $k$th redshift-interval by
\begin{equation}
\begin{split}
	g({\cal{T}}_j, z_k) = \sum_n H(z_k - z_n^{min})\,H(z_n^{max} - z_k) \\ H({\cal{T}}_j - {\cal{T}}_k)C({\cal{T}}_j, z_k) W_n, 
\end{split}
\label{eqgz1}
\end{equation}
where $H$ is the Heaviside step function \footnote{ $ H(x) = 0$ if $x<0$, and $H(x) =1$ for $x\geq0$.  }, $C$ is the completeness fraction (see Equation~\ref{eqgz2}) and the sum extends over all the spectra, $n$, in the sample.  
The $z_n^{min}$ and $z_n^{max}$ are the minimum 
and maximum redshifts at which the redshifted 
\hi\ or OH line can be observed in a spectrum. ${\cal{T}}_k$ represents the 5$\sigma$ integrated 21-cm optical depth threshold in the rest frame 
at $z_k$ = ($\nu_{rest}$/$\nu_k$) - 1, where $\nu_{rest}$ is the rest frequency of the absorption line.  

The spectroscopic redshifts, $z_{spec}$, may not be available for all the targets in a sample. The spectral weight $W_n$ accounts for this possibility.
For targets with known $z_{spec}$, $W_n$=1 always.  For the rest, a value based on the expected redshift-distribution (or photometric redshift) must be assigned.  

In the uGMRT sample, WISE-selected targets account for 16 of the 17 unknown redshifts.  Based on the distribution of targets with 
known $z_{spec}$, we expect 85\% and 70\% of these unknown redshifts to be at $z>0.4$ and $z>0.7$, respectively.  The redshift cuts 
of 0.4 and 0.7 represent the $z_n^{max}$ for \hi\ and OH absorption that can be observed using the L-band (see Equation~\ref{eqgz1}).
We  compute the contribution of targets with unknown redshifts to the sensitivity function by setting $W_n$ = 0.85  and 0.70 for the \hi\ and 
OH searches, respectively. 
This approach minimizes the total redshift path of the survey and leads to a conservative constraint on the incidence of absorbers.

\begin{figure} 
\centerline{\vbox{
\centerline{\hbox{ 
\includegraphics[trim = {0cm 0cm 0cm 0cm}, width=0.5\textwidth,angle=0]{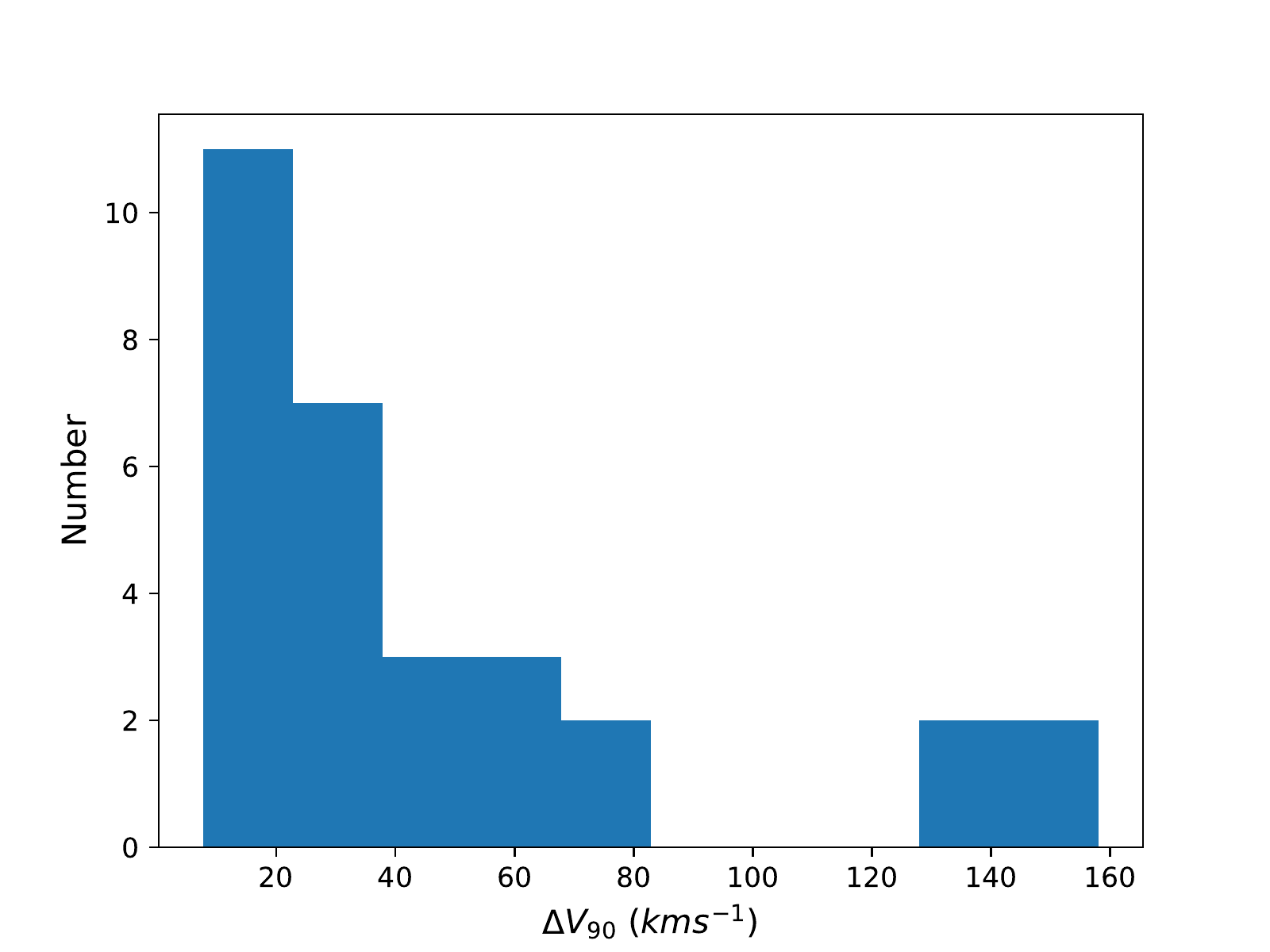}  
}} 
}}  
\vskip+0.0cm  
\caption{
	Distribution of $\Delta V_{90}$ for 30 intervening 21-cm absorbers ($0.02<z_{abs} <3.2$) from our past observations.
} 
\label{fig:dv90}   
\end{figure} 

\begin{figure} 
\centerline{\vbox{
\centerline{\hbox{ 
\includegraphics[trim = {0cm 0cm 0cm 0cm}, width=0.5\textwidth,angle=0]{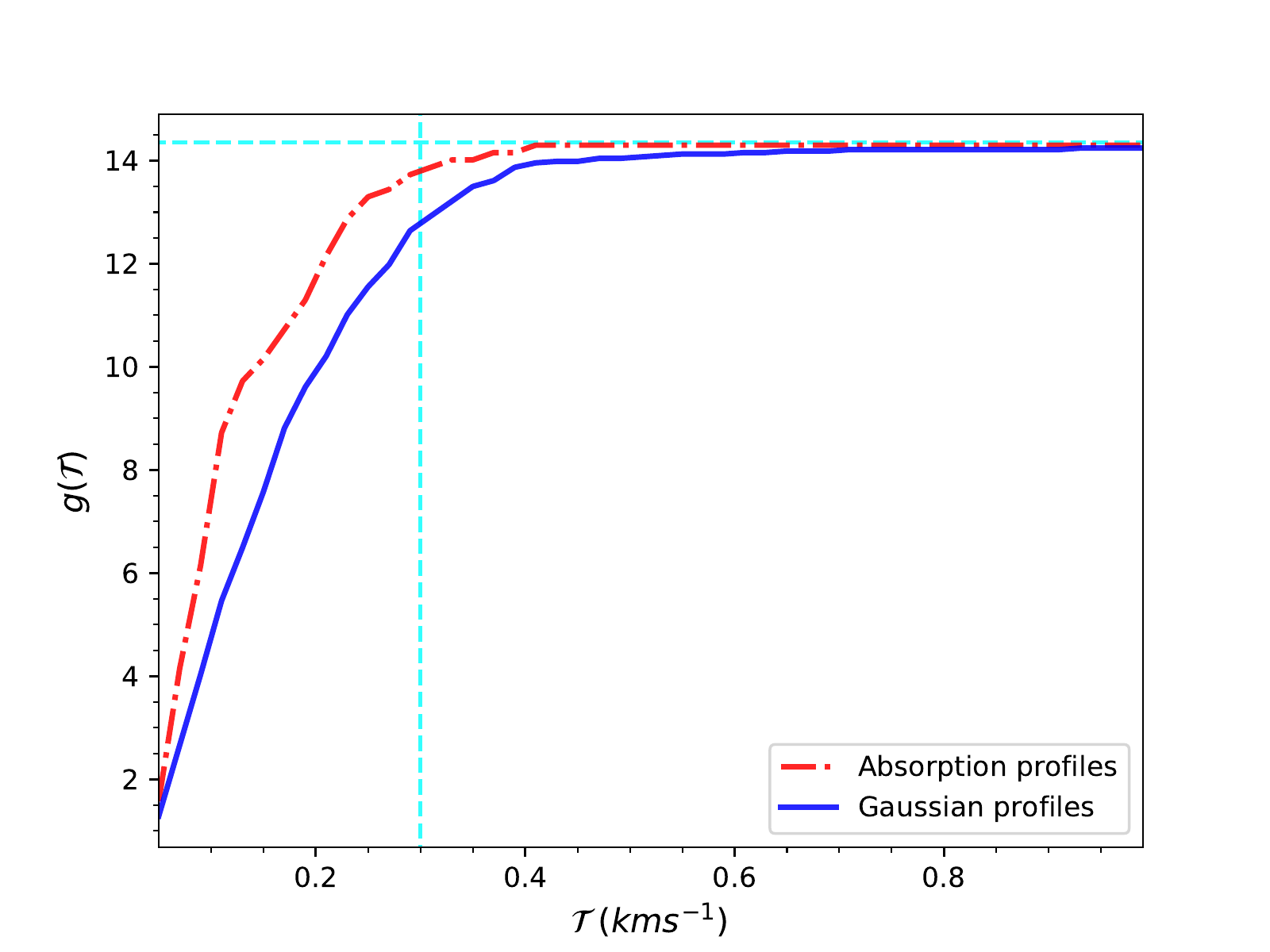}  
}} 
}}  
\vskip+0.0cm  
\caption{
	Completeness corrected total redshift paths ($\Delta z ({\cal{T}}) \equiv g({\cal{T}})$) for the 21-cm line search. 
	The horizontal dashed line represents total redshift path without completeness correction.
	The vertical dashed line corresponds to integrated optical depth, ${\cal{T}}$ = 0.3\,\kms.  
} 
\label{fig:gzcompl}   
\end{figure} 

We compute $g({\cal{T}}, z)$ for 5$\sigma$ detection limits of $N$(\hi)\,=\,$10^{19}$, 5$\times10^{19}$ and $10^{20}$\,\cmsq.  These values adequately cover the 
range of column densities associated with CNM detected in 21-cm absorption surveys of the Milky Way and external galaxies.
For $f_c^{\tiny \hi}$ = 1 and  $T_{\rm s}$ = 100\,K, these column density limits correspond to ${\cal{T}}$ = 0.06, 0.28, and 1.38\,\kms, 
respectively (see Equation~\ref{eq21cm}). 
A particular value of integrated optical depth could either be due to a large peak optical depth, a large velocity spread, or a combination of both. 
In order to quantify the impact of variable optical depth sensitivity across the spectrum, corresponding 
to a given ${\cal{T}}$,  we set up two simulations in which we inject artificial absorption systems at every channel of each sight line and 
apply the same detection algorithm as described in Section~\ref{sec:obs}. 
The output of this exercise is $C({\cal{T}}_j, z_k)$ which we define as
\begin{equation}
\begin{split}
	C({\cal{T}}_j, z_k) = \frac{1}{N_{inj}}  \sum_{i=1}^{N_{inj}}  F({\cal{T}}_j, z_k, \Delta V_{i}), 
\end{split}
\label{eqgz2}
\end{equation}
where $N_{inj}$ is the number of injected systems and $F=1$ if the injected system is detected and 0 if not.

\begin{figure*} 
\centerline{\vbox{
\centerline{\hbox{ 
\includegraphics[trim = {0cm 0cm 0cm 0cm}, width=1.0\textwidth,angle=0]{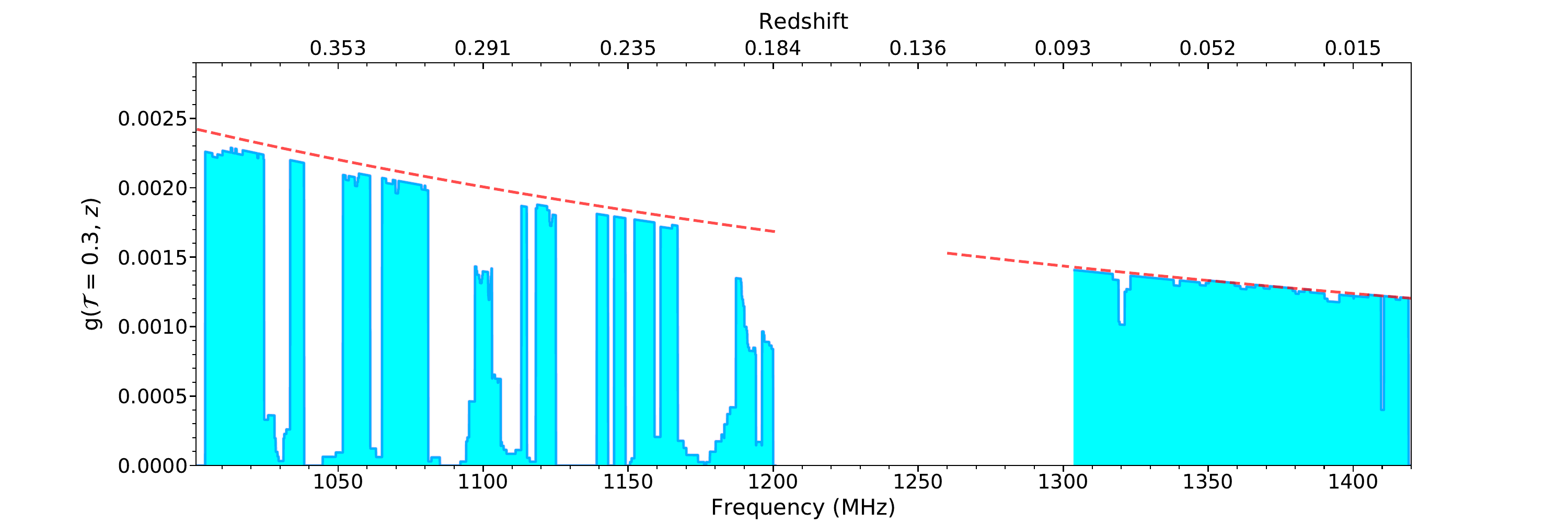}  
}} 
}}  
\vskip+0.0cm  
\caption{
	The sensitivity function, $g(z)$, for \hi\ 21-cm absorbers with integrated 21-cm optical depth ${\cal{T}} \ge$  0.3\,\kms.  
	The dashed line represents $g(z)$ for an idealized survey unaffected by RFI. 
} 
\label{fig:gz}   
\end{figure*} 

First, we consider the profiles of all the intervening absorbers (30 in total) detected from our surveys 
in the last 15 years \citep[][]{Srianand08, Srianand10, Srianand13dib, Gupta09, Gupta10, Gupta12, Gupta13, Gupta18j1243, Dutta16, Dutta17mg2, Dutta17fe2, Dutta17, 
Combes19}. 
The velocity widths ($\Delta V_{90}$) of 
these absorbers, including  the  5th  and 95th percentiles of the apparent optical depth distribution lie in the range of 5-150\,\kms (Fig.~\ref{fig:dv90}). 
Since the uGMRT survey has adequate spectral resolution to resolve these lines, in the first simulation we inject actual profiles of absorbers modeled 
using the detailed Gaussian component fits provided in the above-mentioned references.  
The total or integrated completeness-corrected redshift path of the survey considering all sight lines, which is given by
\begin{equation}
	\Delta z ({\cal{T}}_j) \equiv g({\cal{T}}_j) = \sum_k g({\cal{T}}_j, z_k)\Delta z_k, 
\label{eqgz3}
\end{equation}
from this analysis is shown in Fig.~\ref{fig:gzcompl} as a dash-dotted curve. The total redshift path of 14.3 with no completeness correction  
is plotted as a horizontal dashed line. 

For the second simulation, we model the $\Delta V_{90}$ of absorption profiles shown in Fig.~\ref{fig:dv90} using a log-normal distribution. We inject 200 single Gaussian 
components drawn from this distribution and centered at each pixel to compute $C({\cal{T}}_j, z_k)$.
The $g({\cal{T}}_j)$ corresponding to this is shown as the solid curve in Fig.~\ref{fig:gzcompl}.

Clearly, the completeness-corrected total redshift path drops off rapidly for ${\cal{T}} <$  0.3\,\kms\ (see vertical dashed line in Fig.~\ref{fig:gzcompl}). 
We conclude that the threshold of 0.3\,\kms\ provides the ideal balance between maximizing the search path length and providing 
the ability to detect the column densities of interest. 
Due to the larger number of broader absorption systems injected in the second simulation, the  redshift path in this case (solid curve) falls off more rapidly 
at lower $\cal{T}$. 

In Fig.~\ref{fig:gz}, we show the sensitivity function for ${\cal{T}}$ = 0.3\,\kms. 
The total redshift path for this optical depth sensitivity, considering the conservative estimate from the solid curve in Fig.~\ref{fig:gzcompl}, is  
$\Delta z({\cal{T}}=0.3)=$ 12.9.  
The horizontal dashed lines in Figs.~\ref{fig:lbhigh} and \ref{fig:lblow} 
represent the frequency ranges contributing to the sensitivity function. 
The dips in Fig.~\ref{fig:gz} are primarily caused by the spectral channels removed from the data due to RFI, rather than due to lower signal-to-noise ratio of 
the spectrum.

Next we compute the incidence or number of 21-cm absorbers per unit redshift ($n_{21}$) with integrated optical depth greater than some 
threshold (${\cal{T}}_j$). This quantity is given by 
\begin{equation}
	n_{21} = \sum_{i=1}^{N_{abs}} \frac{1}{\Delta z({\cal{T}}_i)}, 
\label{eqgz4}
\end{equation}
where the sum extends over all the absorbers ($N_{abs}$) with ${\cal{T}}_i \ge {\cal{T}}_j$, and $g({\cal{T}}_i)$ is the redshift path over which the $i$th absorber could be detected.
Over $\Delta z\,({\cal{T}}=0.3)=$ 12.9 with no 21-cm detections and considering the 1$\sigma$ upper limit based on small number Poisson statistics 
\citep[][]{Gehrels86}, we compute $n_{21} <$0.14. 

Another frequently used statistic related to absorption lines is the number of absorbers per comoving path length, $\ell(X)$, which in the current context 
can be defined as
\begin{equation}
	\ell_{21} = \sum_{i=1}^{N_{abs}} \frac{1}{\Delta X({\cal{T}}_i)}, 
\label{eqgz5}
\end{equation}
where the total absorption path length $\Delta X$ is
\begin{equation}
	\Delta X ({\cal{T}}_j)  = \sum_k g({\cal{T}}_j, z_k) \frac{(1 + z)^2}{\sqrt{(\Omega_m(1 + z)^3 + \Omega_\Lambda)}}    \Delta z_k, 
\label{eqgz6}
\end{equation}
For the uGMRT survey, we estimate $\Delta X({\cal{T}}=0.3)=$ 16.9 and the corresponding $\ell_{21}$ = 0.11.

Next, we apply the formalism presented above to the stronger OH main line at 1667.359\,MHz.  
We estimate $n_{\rm OH}$ for threshold ${\cal{T}}_i=0.3$\,\kms.  For $f_c^{\rm OH}$ = 1 and $T_{\rm ex}$ = 3.5\,K (see Section~\ref{sec:pkshighz}),
this corresponds to $N$(OH) = 2.4$\times10^{14}$\,\cmsq\ (see Equation~\ref{eqoh}), which   
 is similar to the $N$(OH)$\sim$10$^{13-14}$\,\cmsq\ observed in diffuse clouds in the Galaxy, but 3$-$550 times lower than the 
column densities of four previously known intervening OH absorbers from gravitational lenses. 
Thus, the threshold adopted for the \hi\ absorption search is also optimal for the OH absorption line search. 
For the latter, we estimate the total redshift path $\Delta z({\cal{T}}=0.3)=$ 15.1 and the number of  
OH absorbers per unit redshift $n_{\rm OH}(z\sim0.40)<$0.12.  The $\Delta X({\cal{T}}=0.3)=$ 24.4 and the corresponding $\ell_{\rm OH}$ = 0.08. 

This is the first time the incidence of OH absorbers has been constrained using a blind survey.  The upper limit is consistent with 
$n_{\rm OH}(z\sim0.1)=0.008^{+0.018}_{-0.008}$ as determined by \citet[][]{Gupta18oh} from a sample of quasar sight lines close to nearby 
galaxies.  The constraint on $n_{21}$ is also a factor of 15 higher than the estimate based on observations 
selectively targeting quasar sight lines with galaxies at low impact parameters \citep[$\rho<$30\,kpc;][]{Dutta16}. 
We discuss the implications of these results in the next section.

\subsection{Expectations from studies based on Mg~{\sc ii} absorption and nearby galaxies}    
\label{sec:mgiigal}  

The \mgiiab\ absorption doublet has long been used as a tracer of metal-enriched gas over a wide range of \hi\ column densities (16 $< $ log\,($N$(\hi)\,\cmsq) $<$ 22) 
in the disks and halos of galaxies.  Its observability from ground-based optical 
telescopes  and the availability of large samples of QSO spectra from SDSS make it suitable for studying the  
CGM-galaxy relationship. 
A well known property of \mgii\ absorbers is that a larger fraction of strong \mgii\ absorbers originate in high-$N$(\hi) gas.
For example, in the HST sample of \citet[][]{Rao06}, there are no DLAs with rest equivalent width $W_r$(\mgiia) $<0.6$\,\AA, and the systems with $W_r$ $>2.0$\,\AA\ always have log\,$N$(\hi)$>$19.0.
This pattern has motivated several searches for \hi\ 21-cm line absorption in strong \mgii\ absorbers, the term that  is often used in the literature to refer to absorbers with 
$W_r$ $>1$\,\AA\ \citep[][]{Prochter06}.

The $n_{21}$-upper limit from the uGMRT survey ($0<z<0.4$) is consistent with the incidences, $n_{21}(z=0.3-0.4) = 0.04^{+0.03}_{-0.02}$ and 
$n_{21}(z=0.5-1.0) = 0.02^{+0.02}_{-0.01}$ estimated by \citet[][]{Dutta17mg2} and \citet[][]{Gupta12}, respectively, 
through observations of strong \mgii\ absorbers.
It is quite likely that surveys based on strong \mgii\ absorbers are biased towards high \hi\ column density systems.  
However, our uGMRT survey, due to its small redshift path, does not have adequate statistical power to address the issues related to \mgii\ selection bias or any redshift evolution.

\begin{figure} 
\centerline{\vbox{
\centerline{\hbox{ 
\includegraphics[trim = {0cm 0cm 0cm 0cm}, width=0.55\textwidth,angle=0]{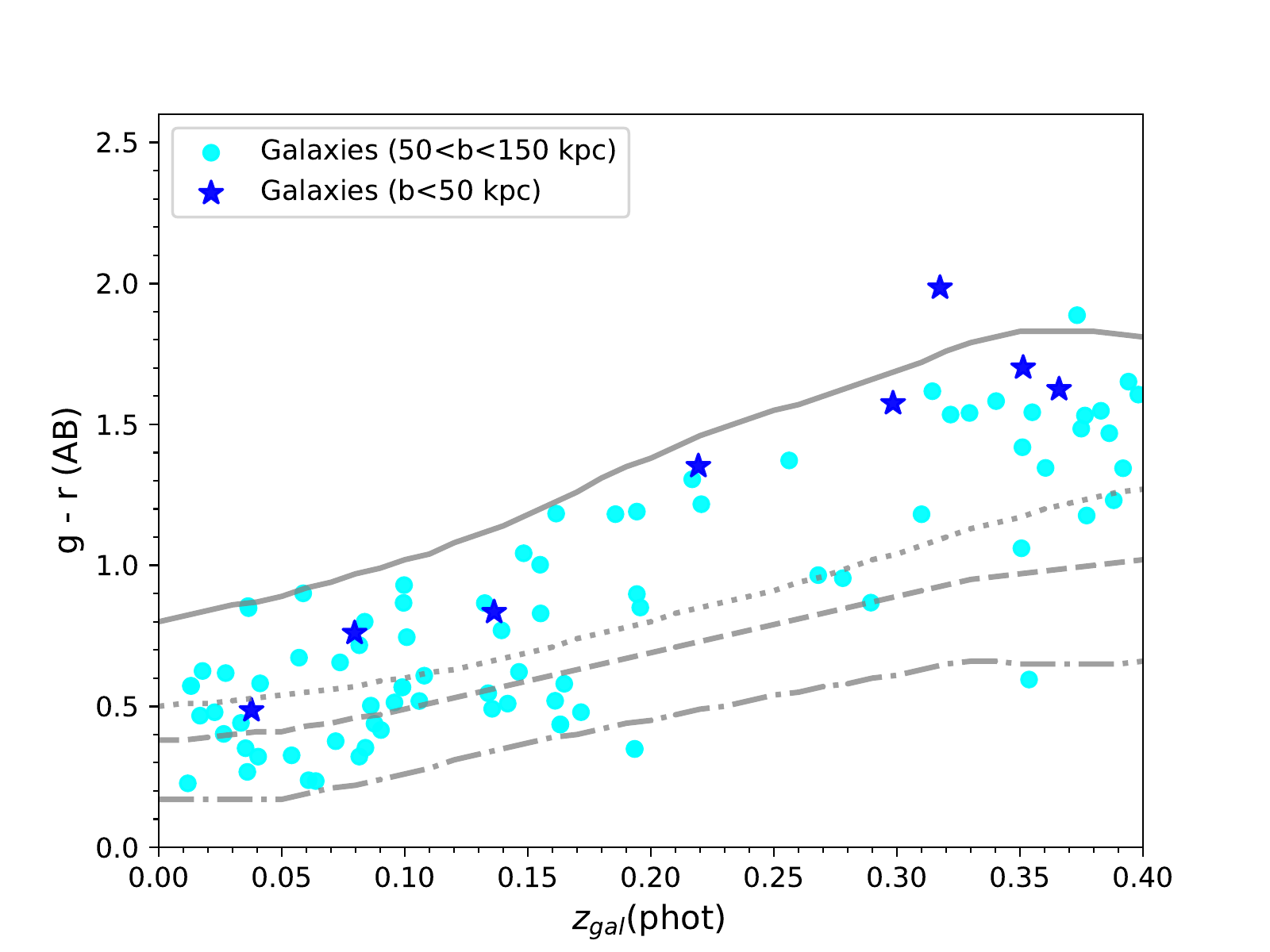}  
}} 
}}  
\vskip+0.0cm  
\caption{
	Observed $g-r$(AB) colors vs. photometric redshifts for galaxies within 50\,kpc ($\bullet$) and 50-150\,kpc ($\bigstar$) of the radio source sight line.
	For reference, colors for SED templates from Coleman are plotted as solid (E), dotted (Sbc), dashed (Scd), and dash-dotted (Im) lines.
} 
\label{fig:galcol}   
\end{figure} 

Still, we cannot rule out the possibility of having sight lines in the sample that pass through luminous galaxies at small impact parameters ($\rho$) but that are not detected 
in 21-cm absorption due to low $N$(\hi), small covering factor of CNM, or a combination of both.
To examine this scenario, we use SDSS photometric redshifts from Data Release 15 to identify 91 galaxies with $r<22$\,mag and within $\rho<$150\,kpc of our sight lines. For this we accept only the most robust photometric 
redshifts with {\tt photoErrorClass = 1}. The observed $g - r$(AB) colors of these galaxies vs. photometric redshifts ($z_{gal}$) are plotted as filled circles and stars in Fig.~\ref{fig:galcol}.    
We also plot colors for four empirically determined SEDs representing E, Sbc, Scd, and Irr galaxies \citep[][]{Coleman80}.  These widely used extended-Coleman SED templates from 
\citet[][]{Bolzonella00} do not take into account the true diversity of galaxy spectral properties, but considering that we are working with photometric redshifts, this broad-brush approach 
has the advantage of simplicity.   

\begin{figure} 
\centerline{\vbox{
\centerline{\hbox{ 
\includegraphics[trim = {0cm 0cm 0cm 0cm}, width=0.55\textwidth,angle=0]{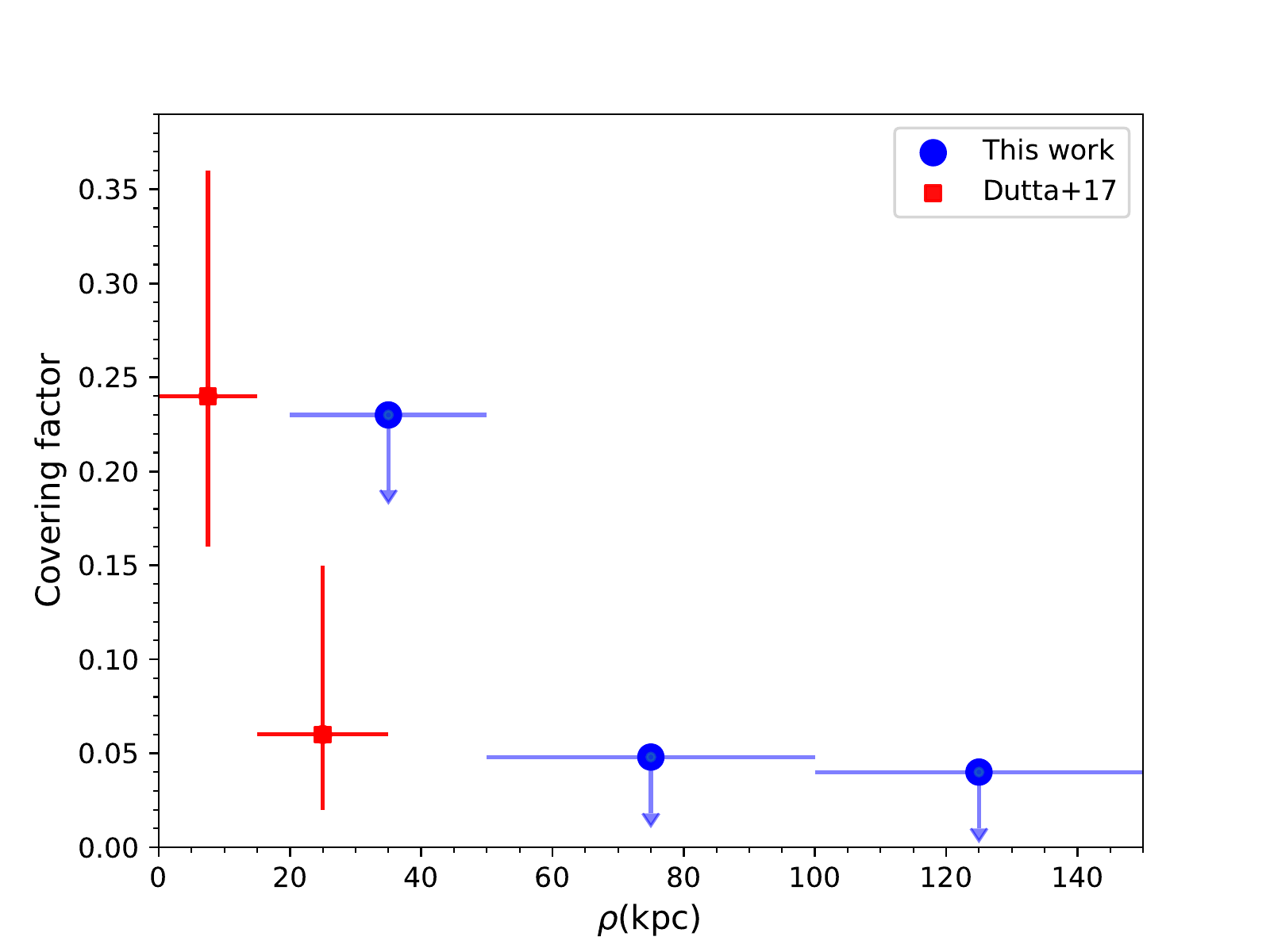}  
}} 
}}  
\vskip+0.0cm  
\caption{
	Covering factor of \hi\ 21-cm absorbers for ${\cal{T}} = 0.3$\,\kms from the uGMRT survey and \citet[][]{Dutta17}.  
} 
\label{fig:c21}   
\end{figure} 

In Fig.~\ref{fig:galcol}, the galaxies with  $\rho<50$\,kpc (eight sight lines) from the uGMRT sample are plotted as stars.  The minimum impact parameter in this subsample is 22\,kpc, and 
the corresponding galaxy has an early type morphology.  
Observations of  nearby galaxies typically detect 21-cm absorption from sight lines passing through optical or extended \hi\ discs of late-type galaxies \citep[][]{Gupta10, Srianand13dib}.  
It is rarer that the absorption is detected from an early type galaxy or tidally disrupted/warped \hi\ disc \citep[][]{Gupta13, Zwaan15}.  
Only one galaxy in the uGMRT subsample has colors similar to those of a late-type galaxy.  But in this case, $\rho\sim$47\,kpc, which is $\sim$3 times 
(i.e., well beyond) the expected size of the \hi\ disc \citep[][]{Noordermeer05}, and $\sim$2 times the largest impact parameter ($\rho$ = 25\,kpc) 
at which 21-cm absorption has been detected in nearby galaxies \citep[][]{Dutta17}. 

Thus, the non-detection of \hi\ and OH absorption from the uGMRT survey can be simply explained by the small redshift path, due to which we could probe only the 
outskirts of star-forming galaxies at $\rho >$30\,kpc.

Although strong \mgii\ absorption is detected at large impact parameters ($\rho\sim150$\,kpc), the covering fraction of the gas is extremely low \citep[$<$0.1;][]{Chen10, Huang16}.
As expected, the colder and higher column density gas detected in \hi\ 21-cm absorption from low-$z$ galaxies is patchier and confined to much smaller impact parameters 
($\rho<$40\,kpc) and the covering fraction falls off from $0.24^{+0.12}_{-0.08}$ at $\rho<$15\,kpc to $0.06^{+0.09}_{-0.04}$ at $\rho$=15$-$35\,kpc \citep[ Fig.~\ref{fig:c21}; see also][]{Dutta17}.  
The 1$\sigma$ upper limits\footnote{Based on small number Poisson statistics \citep[][]{Gehrels86}.} of 0.048 and 0.040 on 21-cm covering factor 
at $50<\rho<100$ and $100<\rho<150$\,kpc, respectively, from our sample are consistent with this picture.  The average value at $50<\rho<150$ is 0.022.

\subsection{Cold gas associated with the AGN}    
\label{sec:assoc}


\begin{deluxetable}{cccc}
	\tablecaption{Upper limits on associated \hi\ and OH absorption. }
\tabletypesize{\scriptsize}
\tablehead{
	\colhead{Source name} & \colhead{$z_{spec}$} & \colhead{$N$(\hi)} & \colhead{$N$(OH)} \\ 
	\vspace{-0.5cm} \\
	\colhead{    } & \colhead{     } & \colhead{ ($10^{19}$\cmsq)    } & \colhead{($10^{14}$\cmsq)} \\ 
	\vspace{-0.5cm} \\
	\colhead{(1) } & \colhead{ (2) } & \colhead{ (3)     } & \colhead{ (4)} \\ 
	\vspace{-0.7cm} \\
} 
\startdata
J0054$-$0333  & 0.210  &   -    &  1.7  \\ 
J0713$+$4349  & 0.518  &   -    &  1.2  \\ 
J0741$+$3112 &   0.632 &   -    &  2.6   \\ 
J0825$+$0309  & 0.506  &   -    &  1.7  \\ 
J0943$-$0819 &   0.228 &  6.4   &  2.4   \\ 
J1035$+$5628 &   0.460 &   -    &  3.3   \\ 
J1156$+$3128  & 0.4171 &   -    &  0.9  \\ 
J2150$+$1449  & 0.4    &  3.8   &   -   \\ 
\label{tab:assoc}
\enddata
\tablecomments{ 
\scriptsize{ Column 1: NVSS ID based on Right Ascension and Declination (J2000). Column 2:  spectroscopic redshift from literature.  
	Columns 3-4: 5$\sigma$ \hi\ and OH upper limits estimated from spectra smoothed by 5 pixels, 
	assuming a line width of 100\,\kms\ and $T_{\rm s}$=100\,K and $T_{\rm s}$=3.5\,K, respectively.}
}
\end{deluxetable}

\hi\ and OH absorption line observations also offer a powerful way to detect cold atomic and molecular gas associated with an AGN \citep[][]{Morganti18}. 
The 21-cm absorption lines associated with AGNs are much broader than those of intervening absorbers. The typical full width at half maximum is 
$\sim$100\,\kms, and $N$(\hi)$>10^{20}$\,\cmsq\ \citep[e.g.,][]{Pihlstrom03, Gupta06}.  OH absorption is relatively less explored but when 
detected the column densities are usually $>$10$^{15}$\cmsq\ \citep[][]{Gupta18oh}.
In Table~\ref{tab:assoc}, we list radio sources for which redshifted \hi\ 21-cm and OH 1667\,MHz line frequencies are unaffected by RFI and lie within the uGMRT band.
\hi\ measurements are possible in two cases and OH in seven cases, but in only one case can we probe both \hi\ and OH.

Interestingly, both the sources with 21-cm absorption measurements show radio spectral energy distributions and morphologies that are consistent with expectations for young 
and evolving radio sources that, depending on jet power and the density of host galaxy ISM, may develop into large scale radio sources \citep[][]{Mantovani98, deVries98}.
For one of them (J0943$-$0819), the host galaxy at $z=0.228$ also has a close companion.
Extremely high detection rates of \hi\ 21-cm absorption have been observed in such sources, suggesting a 
relationship between the presence of cold gas in the circumnuclear region and the fuelling of AGN activity.
Specifically, at a sensitivity of $N$(\hi)$=10^{20}$\,\cmsq, compact young 
radio sources exhibit detection rates of 30-45\%, and those associated with mergers have a detection rate of $\sim$80\% \citep[][]{Gupta06, Dutta19}. 
Even though excellent column density sensitivity is achieved in our observations, no 21-cm absorption is detected.  
The 5$\sigma$ upper limits on \hi\ column densities estimated assuming a line width of 100\,\kms\ are in the range of (4-7)$\times10^{19}$\,\cmsq.  
The sample size is too small for us to draw any general conclusions.  

For OH, the upper limits on column densities are in the range of (1-3)$\times10^{14}$\,\cmsq (Table~\ref{tab:assoc}).   
As previously mentioned, OH absorption associated with powerful AGNs is much less explored.  OH studies have generally focused on starburst 
galaxies or late and early-type galaxies with large reservoirs of dense molecular gas already confirmed through the presence of emission lines of 
CO, HCN, etc. \citep[e.g.,][]{Baan89, McBride15}.
None of the radio sources in the sample here possess such properties.   

Larger samples with measurements of both \hi\ and OH absorption from upcoming SKA pathfinder and precursor surveys will help constrain 
the presence of cold gas and test models relating the presence of cold gas to the orientation-based unified scheme or the evolutionary stage 
of the AGN \citep[][]{Gupta06uni, Maccagni17}.

\section{Summary and Outlook}    
\label{sec:summ}  

This paper describes the current capabilities of {\tt ARTIP},  which are being developed specifically to process radio interferometric data for MALS. 
{\tt ARTIP} is extremely portable and general enough to process datasets at cm wavelengths from any radio interferometer, e.g., uGMRT, VLA, and MeerKAT \citep[see for example][]{Gupta18j1243, Gupta18oh, Combes19}.
It is written in Python 3.6, and utilizes the CASA tools and task framework to perform flagging, calibration, and imaging. 
The pipeline can be deployed as a single node application or as a multinode application on a cluster.
The complex workflows, related to the spectral line processing of wideband datasets on a cluster, can be set up, triggered, and monitored 
through an intuitive user interface.  The interface is designed to enable the geographically-distributed MALS team to collaboratively 
set up and trigger {\tt ARTIP} execution, monitor progress, and manage data products. 

We have illustrated the spectral line capabilities of {\tt ARTIP} by applying it to the observations (date: December 19, 2019) of a field centered on PKS\,1830-211. 
This is the first MALS observation using the full MeerKAT array and 32K channel mode of the SKARAB correlator.  
The wideband continuum image of the target has a dynamic range of $\sim$78500.
For $T_{\rm s}$ = 100\,K, $T_{\rm ex}$ = 3.5\,K, unit covering factor and line width of 6\,\kms, the spectrum has the sensitivity to detect gas with  
column densities of $N$(\hi) $>$ 1.4$\times10^{18}$\,\cmsq\ and $N$(OH) $>$ 5.9$\times10^{12}$\,\cmsq at 5$\sigma$ significance, respectively.  
With merely 40\,minutes of on-source time, this is the most sensitive spectrum of the target to date. 

We detect the known \hi\ 21-cm absorption ($z=0.19$) at $\sim$1191\,MHz, and 
OH 1665 and 1667\,MHz absorption ($z=0.89$) at $\sim$883.5 and 884.5\,MHz respectively.  
We measure $N$(\hi) = $(2.77 \pm 0.04)\times10^{20}$ (${T_{\rm s}}/{100\,{\rm K}}$)(${0.5}/{f_{\rm c}^{\tiny \hi}}$)\,\cmsq\ and 
$N$(OH) = (1.46  $\pm$ 0.05)$\times$10$^{15}$(${T_{\rm ex}}/{5.14\,{\rm K}}$)(${1.0}/{f_c^{\rm OH}}$)\,\cmsq.  
Both $f_{\rm c}^{\tiny \hi}$ and $f_c^{\rm OH}$ are less than unity, and most likely $f_c^{\rm OH}$ $\le$ $f_{\rm c}^{\tiny \hi}$.  
So the column density estimates are strictly lower limits.

The $z=0.89$ absorber is particularly special: at high frequencies, numerous molecular species, such as CO, HCO$^+$, HCN, HNC, and their rare 
isotopologues, have been detected towards the NE and SW continuum peaks \citep[e.g.,][]{Wiklind98, Muller14}. 
The MeerKAT spectrum has a total bandwidth of 856\,MHz and  covers several transitions of NS, CCS, and CH$_2$CN for the $z=0.89$ absorber.  
We provide upper limits on the column densities of these species.

\begin{deluxetable}{cccc}
\tabletypesize{\small}
\tablecaption{Summary of recent blind 21-cm line searches.}
\tablehead{
\colhead{Survey} &  \colhead{Reference      } & \colhead{$z$-range}     & \colhead{$\Delta z^\dag$} \\
}
\startdata
        uGMRT L-band          &  This work                &   0 - 0.4   &   12.9   \\
	ASKAP FLASH           &  \citet[][]{Allison20}    & 0.34 - 0.79 &    3.7$^\ast$   \\
	ALFALFA$^\ddag$       &  \citet[][]{Darling11}    &   0 - 0.05  &    7.0   \\
\label{tab:blind}
\enddata
\tablecomments{ $\dag$: Total redshift path for DLAs i.e.\ $N$(\hi)$>2\times10^{20}$\,\cmsq.  $\ddag$: The Arecibo Legacy Fast Arecibo 
L-Band Feed Array (ALFALFA) Survey. $\ast$: Based on early science survey of the GAMA 23 field. For rarer super-DLAs ($N$(\hi)$\ge 2\times 10^{21}$\,\cmsq), $\Delta z\sim$63. 
	}
\end{deluxetable}

We demonstrate the versatility of ARTIP in handling datasets involving realistic observing scenarios by applying it to a sample of bright radio sources observed 
with the  uGMRT to carry out a blind search for \hi\ and OH absorbers at $z<0.4$ and $z<0.7$, respectively.  
From the uGMRT survey, we estimate the numbers of \hi\ and OH absorbers per unit redshift to be $n_{21}(z\sim0.18)<$0.14 and $n_{\rm OH}(z\sim0.40)<$0.12, respectively. 
These have been estimated for 5$\sigma$ column density thresholds of $N$(\hi) = $5 \times 10^{19}$\,\cmsq (spin temperature = 100\,K) and 
$N$(OH) = $2.4 \times 10^{14}$\,\cmsq (excitation temperature = 3.5\,K). 
This is the first time the incidence of OH absorbers has been constrained using a blind survey. 

The redshift paths of recent blind 21-cm line surveys to detect CNM in gas with $N$(\hi)$\ge 2\times 10^{20}$\,\cmsq\ are summarized in Table~\ref{tab:blind}.
Thanks to the excellent optical depth sensitivity of the uGMRT survey, the redshift path to detect CNM in DLAs is now increased by a factor of 2. Despite this,
no associated or intervening \hi\ absorption is detected.
We show that the non-detection of \hi\ and OH absorbers, and upper limits on $n_{21}$ and $\rm n_{OH}$, can be explained by the small redshift path of the 
survey ($\Delta$$z\sim$15) and the lack of star-forming galaxies at impact parameter $\rho<30$\,kpc from the line of sight to the background radio sources. 
The survey has allowed us to constrain the cold gas covering factor of galaxies at large impact parameters (50\,kpc$<\rho<$150\,kpc) to be less than 0.022. 
In the near future, MALS will substantially increase ($\Delta$$z\sim10^{3-4}$) the redshift path and provide stringent constraints on the cold gas fraction of 
galaxies in diverse environments.

The proper science observations of MALS using the L-band receiver have started. Each target at L-band will be observed for $\sim$56\,minutes.  
The first of the MALS observations were carried out on June 14, 2020.  
Prior to this and while this paper was being written, we also observed the quasar J163956.35+112758.7 as a 
science-verification target for 56\,minutes, to mimic a MALS observation exactly (see Section~\ref{sec:appj1639} for details).  
These observation deliver continuum image and the spectrum with rms consistent with theoretical thermal noise. 

It is only natural that as we process the upcoming MeerKAT datasets, we will uncover new challenges and some of the approaches presented here will need to be updated.  
In the meantime, development is underway to incorporate direction-dependent imaging corrections via the {\tt AW-Projection} algorithm (Sekhar et al. in preparation) to the {\tt ARTIP-CONT} pipeline and improve the flagging of low-level RFI.  
These capabilities will mitigate the artefacts associated with the bright off-axis sources seen in Fig.~\ref{fig:j1639}. 
These updates to the pipeline and the details of radio continuum and polarization processing will be detailed in future publications. 

\acknowledgments

NG, PN, PPJ and RS acknowledge support from the Indo-French Centre for the Promotion of Advanced Research 
(Centre Franco-Indien pour la promotion de la recherche avanc\'ee) under Project {\tt 5504-B}.
PK is partially supported by the BMBF project 05A17PC2for D-MeerKAT.
This research was supported by the Munich Institute for Astro- and Particle Physics (MIAPP), which is funded by 
the Deutsche Forschungsgemeinschaft (DFG, German Research Foundation) under Germany´s Excellence Strategy – EXC-2094 – 390783311.
VPK acknowledges partial support from  NASA grant NNX17AJ126G.
The MeerKAT telescope is operated by the South African Radio Astronomy Observatory, which is a facility of 
the National Research Foundation, an agency of the Department of Science and Innovation.
The GMRT is run by the National Centre for Radio Astrophysics of the Tata Institute of Fundamental Research.
The Common Astronomy Software Applications (CASA) package is 
developed by an international consortium of scientists based at the National Radio 
Astronomical Observatory (NRAO), the European Southern Observatory (ESO), the National 
Astronomical Observatory of Japan (NAOJ), the Academia Sinica Institute of Astronomy 
and Astrophysics (ASIAA), the CSIRO division for Astronomy and Space Science (CASS), 
and the Netherlands Institute for Radio Astronomy (ASTRON) under the guidance of NRAO.
The National Radio Astronomy Observatory is a facility of the National Science Foundation 
operated under cooperative 
agreement by Associated Universities, Inc.
We acknowledge using the 
Inter-University Institute for Data Intensive Astronomy (IDIA) infrastructure for transferring MALS science verification datasets to IUCAA. 
IDIA is a partnership of three South African universities: the University of Cape Town, the University of the Western Cape and the University of Pretoria.  

\facilities{MeerKAT, uGMRT}

\software{ARTIP, Astropy \citep{astropy:2013, astropy:2018}, CASA \citep{McMullin2007} and Matplotlib \citep{Hunter2007}.}

\appendix
\counterwithin{figure}{section}

\section{Details of ARTIP }
\label{sec:artipdetails}

\subsection{Design, capabilities and deployment}
\label{sec:artipdesign}
%
{\tt ARTIP} follows a stage driven architecture, in which outputs from previous stages are used by subsequent stages.    
The pipeline design is modular and ensures the flexibility and extensibility of various processing steps.  Specifically, each processing stage is not 
only fully configurable through a {\tt JavaScript Object Notation} ({\tt JSON}) or {\tt YAML} file, but also replaceable by external software.  
This flexibility is essential, as various 
algorithmic choices for full Stokes wideband calibration and imaging are under active development and testing. 

At the highest level, {\tt ARTIP} is split into the following four components:  
\begin{enumerate}
	\item{\tt ARTIP-CAL:} It ingests a measurement set (ms) containing raw visibilities, and produces as output calibrated target source visibilities. 
	Measurement sets containing multiple target sources and calibrators can be handled.  Only calibrator data are flagged, but 
	these flags can be logically extended to target sources. 
\item{\tt ARTIP-CUBE:} It takes the output of {\tt CAL} and performs spectral-line imaging. The implementation includes RFI flagging, 
	self-calibration, continuum subtraction, and spectral line (i.e., cube) imaging.
\item{\tt ARTIP-CONT:} It takes the calibrated measurement set from {\tt CAL} and generates wideband radio continuum images.  
	The data can be flagged and self-calibrated as needed.
\item{\tt ARTIP-Diagnostic:} It operates on the outputs of {\tt CAL}, {\tt CONT} and {\tt CUBE} to generate diagnostic plots and statistics.

\end{enumerate}
%
These are Python packages written in {\tt Python\,3.6} using standard modules and libraries such as {\tt Numpy} and {\tt Pandas}. 
They make extensive use of {\tt CASA} tasks and tools, and custom codes for data manipulation, flagging, and diagnostics, and for parallelization of various steps.  
In the current version, the calibration and imaging steps are carried out exclusively using {\tt CASA}.

{\tt ARTIP} can be deployed as a single node application on a workstation or as a multinode application on a cluster. The pipeline stages can then be configured and triggered using 
command line arguments to process the data. However, the full capabilities of {\tt ARTIP} are realised only when it is deployed along with its 
{\tt Pipeline Manager} on a cluster supporting parallel computing.   

The {\tt Pipeline Manager} is a web application.   
It features an intuitive and user friendly interface for collaborative {\tt ARTIP} execution, data partitioning strategy control, progress monitoring, 
and management of {\tt pipeline products}. 
The {\tt Pipeline Manager} service can be launched either {\it (i)} via the command line where it operates in a standalone environment, or 
{\it (ii)} using uWSGI server in a production-like environment, as in Fig.~\ref{fig:artipscm}, for handling multiple parallel requests. 

In Fig.~\ref{fig:artipscm}, we presented the complete deployment of {\tt ARTIP} in the MALS A3, i.e., {\it Automated Processing, Analysis, and Archiving} 
environment. The MALS A3-Server is a microservice architecture based implementation that hosts an API and user interface (UI) server, and provides authorized access to the {\tt Pipeline Manager}. 
It also hosts the MALS database (DB) implemented in {\tt PostgreSQL}.  The DB contains 
a list of all the potential MALS pointings and external links to various multi-wavelength surveys such as 
the NRAO VLA Sky Survey \citep[NVSS;][]{Condon98}, 
the Sydney University Molonglo Sky Survey \citep[SUMSS;][]{Mauch03}, 
the Panoramic Survey Telescope and Rapid Response System \citep[PanSTARRS;][]{Chambers16}, 
the Sloan Digital Sky Survey \citep[SDSS;][]{York00}, and 
the Wide-field Infrared Survey Explorer \citep[WISE;][]{Wright10} All-sky data release\footnote{The information from more surveys will be added as needed.}.  

The DB also contains the observing and processing states of each potential pointing. After successful MeerKAT observations and data processing, 
the observing and processing logs, 
the configuration files and the diagnostic plots also get deposited to the DB.  The authenticated MALS-users use these to perform quality checks and 
identify datasets to be processed on the MALS processing cluster, i.e., {\tt VROOM} at the Inter-University Centre for Astronomy and 
Astrophysics (IUCAA) in India. 

The specifications of {\tt VROOM} are provided in Fig.~\ref{fig:artipscm}. It is set up to support 4\,PB of fast 
{\tt DataDirect Networks} ({\tt DDN}) storage for processing, of which 2.5\,PB are already available.
On {\tt VROOM}, {\tt ARTIP} is integrated with job scheduling software {\tt Portable Batch System (PBS)}, and using the {\tt Pipeline Manager} a MALS user can:
\begin{enumerate}
	\item Set up configurations for {\tt CAL}, {\tt CUBE}, {\tt CONT}, and {\tt Diagnostic}.
	\item Trigger {\tt ARTIP} on the cluster, and monitor progress through traffic light status 
		(Grey: {\it not started}; Yellow: {\it in progress}; Blue: {\it finished};    
		Red: {\it failed}) and logs that are continuously updated.
	\item Reset {\tt ARTIP} to a particular stage and restart the processing.  The {\tt pipeline products} including data editing flags are appropriately adjusted.
	\item Selectively push the diagnostic plots and data products to the DB for quality control and, if desired, to the 
		{\tt medium term storage} in the MALS repository for archiving.  	
\end{enumerate}
In the MALS repository, the {\tt medium term storage} is made up of disks storing (i) raw data, which are ready for processing, 
and (ii) {\tt pipeline products}, which may be eventually moved to tapes in the {\tt long term storage}. The maximum capacity of the {\tt long term storage} 
is estimated to be 8\,PB.

\subsection{Processing stages}
\label{sec:artipstages}
%

\begin{figure*} 
\centerline{\vbox{
\centerline{\hbox{ 
\includegraphics[trim = {0cm 0.0cm 0cm 0cm}, width=0.9\textwidth,angle=0]{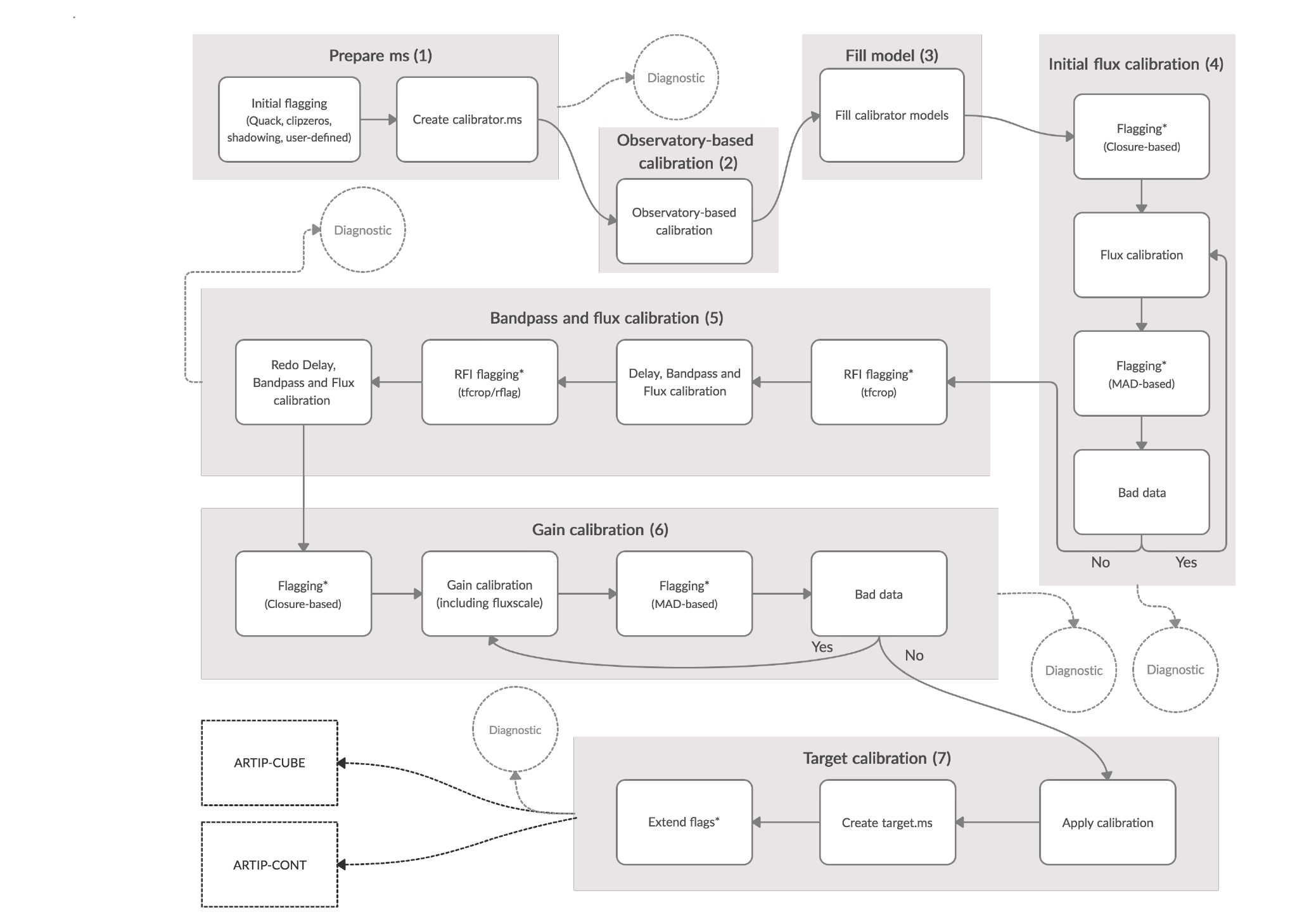}  
}} 
}}  
\vskip+0.0cm  
	\caption{ Schematic of {\tt ARTIP-CAL} processing stages. The stages are numbered (1)$-$(7) for quick referencing.
        The $\ast$ indicates that the substage is optional. 
	The {\tt Diagnostic} represented by dashed lines may be triggered at the end of each stage or after all the stages have completed.  } 
\label{fig:artipcal}   
\end{figure*} 

\subsubsection{ARTIP-CAL} 
\label{sec:cal}

The schematic of {\tt ARTIP-CAL} for calibration of parallel hand visibilities is presented in Fig.~\ref{fig:artipcal}.  
The details of polarization calibration, which is yet to be implemented, will be presented in a future paper.
In total there are seven stages that can be toggled on or off as needed.  These stages, which are implemented as {\tt Python} classes, are further split into substages, which are either subclasses or functions.  The substages are fully configurable to control the desired level of flagging and parameters for calibration. The calibration tables are always regenerated after flagging.   

Stage (1) is called {\tt Prepare ms}. Prior to this stage, original ms is copied if it is not already present, and then 
observatory-based flags are applied. The latter consist of: {\it (i)} applying RFI masks to eliminate persistent interference, 
{\it (ii)} quacking i.e., clipping data at scan boundaries and spectral-window edges, and 
{\it (iii)} flagging data corresponding to non-working antennas, shadowing, zeros, and NaNs. 
The output of this stage is an ms containing only the calibrator data ({\it calibrator.ms}).  Stages (2)$-$(6) operate on {\it calibrator.ms}.

\begin{figure*} 
\centerline{\vbox{
\centerline{\hbox{ 
\includegraphics[trim = {2.5cm 4.5cm 0cm 4.5cm}, width=1.05\textwidth,angle=0]{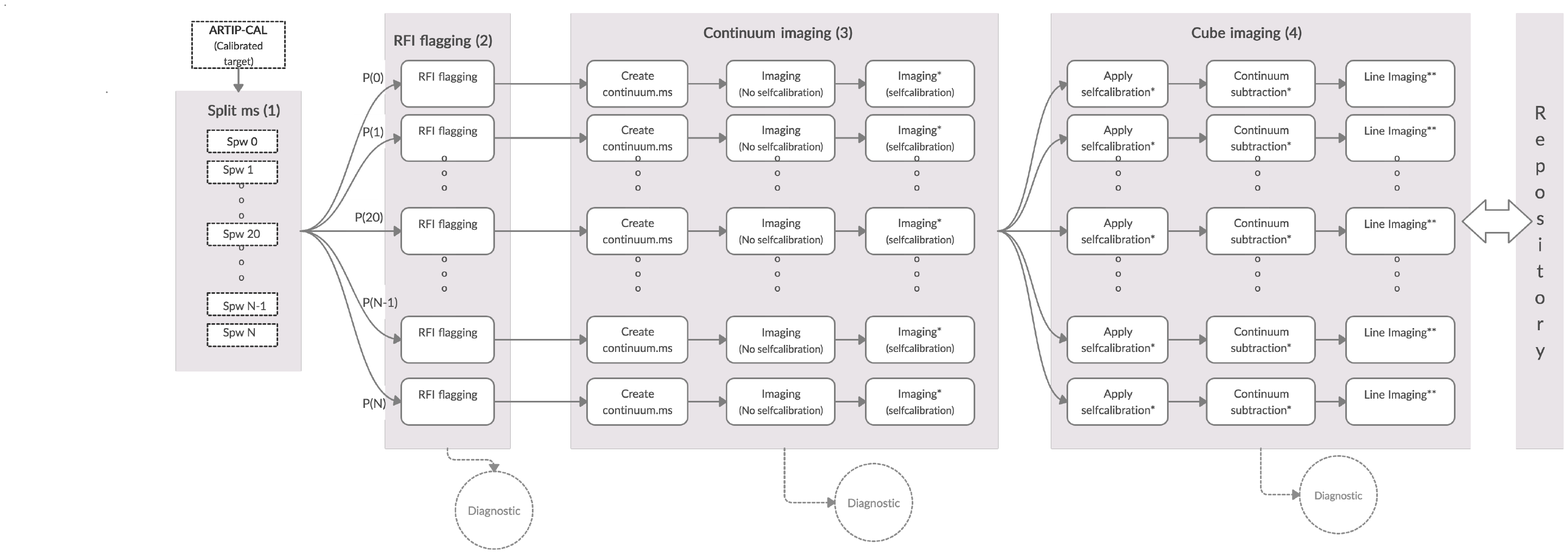}  
}} 
}}  
\vskip+0.0cm  
	\caption{ Schematic of {\tt ARTIP-CUBE} processing stages. P(1), P(2), ...., P$_n$ are processes to $n$ worker nodes. 
	The stages are numbered (1)$-$(4) for quick referencing. The $\ast$ indicates that the substage is optional. 
	The $\ast\ast$ indicates that the substage may be run any number of times to generate images of different specifications.
	The {\tt Diagnostic} represented by dashed lines may be triggered at the end of each stage or after all the stages have completed.   
	} 
\label{fig:artipcube}   
\end{figure*} 

Stage (2) is responsible for generating calibration tables to correct for atmospheric optical depth, ionospheric effects, and variations in gains of 
antennas and electronics. The sequence of steps executed here is specific to the telescope and observing frequency; hence, this stage is called 
{\tt Observatory-based calibration}. This stage is optional, but if it is executed, the output calibration tables are utilized in subsequent stages.

In stage (3), the model visibilities for the primary flux density calibrators are filled in {\it calibrator.ms}.  There is an option to specify clean 
component images for models unavailable in CASA. 

The objective of stage (4) is only to use flux density calibrators to identify  data that are bad for all spectral windows, for example, due to non-working antennas, 
bad baselines or time ranges. The approach to flagging is hierarchical, i.e., instead of chasing individual outliers, the focus is on first identifying 
antennas and then baselines that are generally under-performing for some or most of the time.  The flags generated from stage (4) may be extended 
to other sources in the ms, but the calibration tables are not used hereafter.

In stage (4), raw visibilities from a single or a small range of user-specified (and assumed to be RFI-free) frequency channels 
are first analyzed for phase dispersion and closures. Antennas, baselines, or time ranges with values beyond the user-defined thresholds are flagged.  The flux density calibrator 
data are then calibrated and further examined using  median and median absolute deviation (MAD) statistics to identify bad time ranges (i) across all antennas, 
(ii) for a particular antenna, and (iii) for a particular baseline.  This process is repeated until no bad data in excess of a user-defined threshold are found. 
Obviously, this step will converge faster if the bulk of the bad data are already flagged (e.g., using information from observing logs) in stage (1).

Stage (5) performs delay, bandpass, and flux density calibration, with RFI flagging. 
The RFI flagging of the bandpass calibrator is split into two parts: (i) use {\tt tfcrop} to flag strong RFI prior to calibration, and 
(ii) use {\tt tfcrop} and {\tt rflag} for detailed RFI flagging of the calibrated data. The normal procedure is to  
flag strong RFI, then perform and apply delay, bandpass, and flux density calibration, and then do the detailed RFI flagging.  
All the calibration tables are regenerated after this sequence.  
The inputs for {\tt tfcrop} and {\tt rflag} are configurable for arbitrarily defined frequency ranges with independent flagging thresholds. 

In stage (6), gains for secondary calibrators are determined.  Flagging options based on phase dispersion and closure (see stage (4)) 
are available.  After flagging, the calibration tables are regenerated, and the flux density scale is bootstrapped from primary calibrators.  

Finally, in stage (7), all the calibrations determined so far are applied to the target sources(s), which are then split into separate measurement sets, 
i.e., one ms per source.  
Optionally, the flags determined so far can be extended to target sources using simple logic such as flag data for scan no.\ {\tt n} in original ms 
if the data for scans {\tt n-1} and {\tt n+1} in {\it calibrator.ms} are fully flagged.  
  
Through {\tt Pipeline Manager} a user can monitor the progress and trigger {\tt Diagnostic} at the end of each stage as shown in Fig.~\ref{fig:artipcal}, 
or once after stage (7), to examine the quality of 
calibration and flagging.  The diagnostics may include plots of calibration tables, calibrated visibilities, and flagging statistics per stage. As mentioned previously, 
if needed, the processing may be reset to any previous stage and then restarted.
%

\subsubsection{ARTIP-CUBE}      
\label{sec:cube}

The calibrated target source data from {\tt CAL} are used for spectral line imaging. Thus, the input to {\tt CUBE} is an ms with a single target source. 
This ms may already have multiple spectral windows, or the user can define a set of overlapping spectral windows.
In stage (1), the data corresponding to spectral windows are immediately partitioned 
into distinct measurement sets.  These individual datasets are then processed through the subsequent stages as shown in Fig.~\ref{fig:artipcube}.  The processing of these  spectral windows is then distributed in an embarrassingly  parallel manner across the compute nodes using the {\tt PBS}.
The {\tt Pipeline Manager} provides complete flexibility in the selection of spectral windows to be processed up to which stage(s).
This approach makes the processing robust. In case a particular job for a spectral window fails, then it can be resumed without affecting the 
processing of other jobs.

In stage (2), the dataset for each spectral window is independently flagged for RFI using {\tt rflag} and {\tt tfcrop}. 
In stage (3), each spectral window is averaged in frequency to generate continuum datasets that can be optionally flagged using methods 
implemented in stage (4) of {\tt CAL}.  Each spectral window is then imaged and self-calibrated independently. 
For this step a user may specify the number of phase-only and amplitude-and-phase self-calibration loops to be performed.
If required, it is possible to apply calibration solutions from one spectral window to any other spectral window(s). 
Considering that some spectral windows are more affected by RFI than others, this approach is not only important for obtaining the best output for each spectral 
window, but also makes the processing more robust and fault tolerant.  
In stage (4), self-calibration solutions obtained from the previous stage are applied to RFI-flagged line datasets from stage (2). Then, 
continuum subtraction and line imaging are performed. Here also the same flexibility in applying calibration solutions across spectral windows is available. 

In stage (4), the continuum subtraction can be performed either by (i) subtracting model visibilities determined from a continuum source model, i.e., the {\tt uvsub} approach, or (ii) fitting polynomials to real and complex parts of each visibility spectrum, i.e., the {\tt uvlin} 
(also known as {\tt uvcontsub}) approach {\bf \citep[][]{Cornwell92} }.  For the {\tt uvsub} approach, the default option is to use the continuum model from the same spectral window;  
however it is possible to specify a model from an external process.  For example, a full primary beam wideband model generated using {\tt CONT} with 
an appropriate number of Taylor coefficients to predict the correct spectral behaviour may be specified \citep[][]{Rau11}. 

It is important to note that the last substage, {\tt Line Imaging},  may be run any number of times to generate images of different specifications which can 
then be archived under different contexts e.g.\ a low spatial and spectral resolution image for \hi\ emission and a high resolution image for \hi\ absorption.

In {\tt CUBE}, the {\tt Pipeline Manager} allows users to track the processing of each spectral window.  The diagnostics consist of plots of calibration tables, 
flagging statistics, details of the continuum image (peak flux, synthesized beam, rms, etc.) and the continuum subtracted spectrum of the strongest source in the 
continuum image.
If needed, the processing for selected spectral windows may be reset to a previous stage and then restarted.

\subsubsection{ARTIP-CONT}      
\label{sec:cont}

In brief, it is implemented as a special case of {\tt CUBE} where only stages-(1) to -(3) are executed.
Here, the input ms from {\tt CAL} is not split into separate ms per spectral window. The overall processing strategy is described in Section~\ref{sec:pks}.
The last substage of {\tt Continuum imaging}, i.e., stage (3) may be executed any number of times to generate continuum images of different specifications using the same self-calibrated visibilities.

\subsubsection{ARTIP-Diagnostic} 
\label{sec:cal}

The {\tt Diagnostic} uses outputs of {\tt CAL},  {\tt CUBE}, and {\tt CONT}, and the metadata stored as part of the measurement set, to generate various 
diagnostic plots and statistics that can be viewed through the {\tt Pipeline Manager}. The main objective is to allow assessment of the quality of 
flagging and calibration through various stages of the pipeline.
It essentially follows the same stage driven architecture with one-to-one mapping with respect to the processing stages.
The types of diagnostics generated for {\tt CAL} and {\tt CUBE} are briefly mentioned in previous sections, and 
alluded to in Section~\ref{sec:pks}.  Further details of this package are beyond the scope of this paper.  

\section{MALS: J1639+1127}
\label{sec:appj1639}

The quasar J163956.35+112758.7 was observed on April 1, 2020.   The MeerKAT dataset was processed using the same methods as previously described for PKS\,1830-211.
In Fig.~\ref{fig:j1639} we present the wideband continuum image centered on the quasar J1639+1127, which has a flux density of 178.5 $\pm$ 0.8\,mJy at 1270\,MHz. 
The in-band integrated spectral index, $\alpha$ = -0.08.
The known \hi\ 21-cm absorption line at $z=0.079098$ detected towards 
the quasar is also shown in Fig.~\ref{fig:j1639}.  
The continuum image ({\tt robust=0}) has an rms of {\bf $\sim$15$\mu$Jy\,beam$^{-1}$} and imaging dynamic range of $\sim$11000. 
The rms in the unsmoothed spectrum with {\tt robust=2}, i.e., natural weighting, is 0.51\,mJy\,beam$^{-1}$. 
This value is actually slightly better than the expected theoretical thermal noise (0.59\,mJy\,beam$^{-1}$). 
The integrated optical depth and line width also show excellent consistency with the GMRT measurement (see Fig.~\ref{fig:j1639}).

\setcounter{figure}{0}    

\begin{figure} 
\vspace{1.0cm}
\centerline{\vbox{
\centerline{\hbox{
\includegraphics[trim = {0.0cm 11.5cm 4.0cm 3.0cm}, width=0.50\textwidth,angle=0]{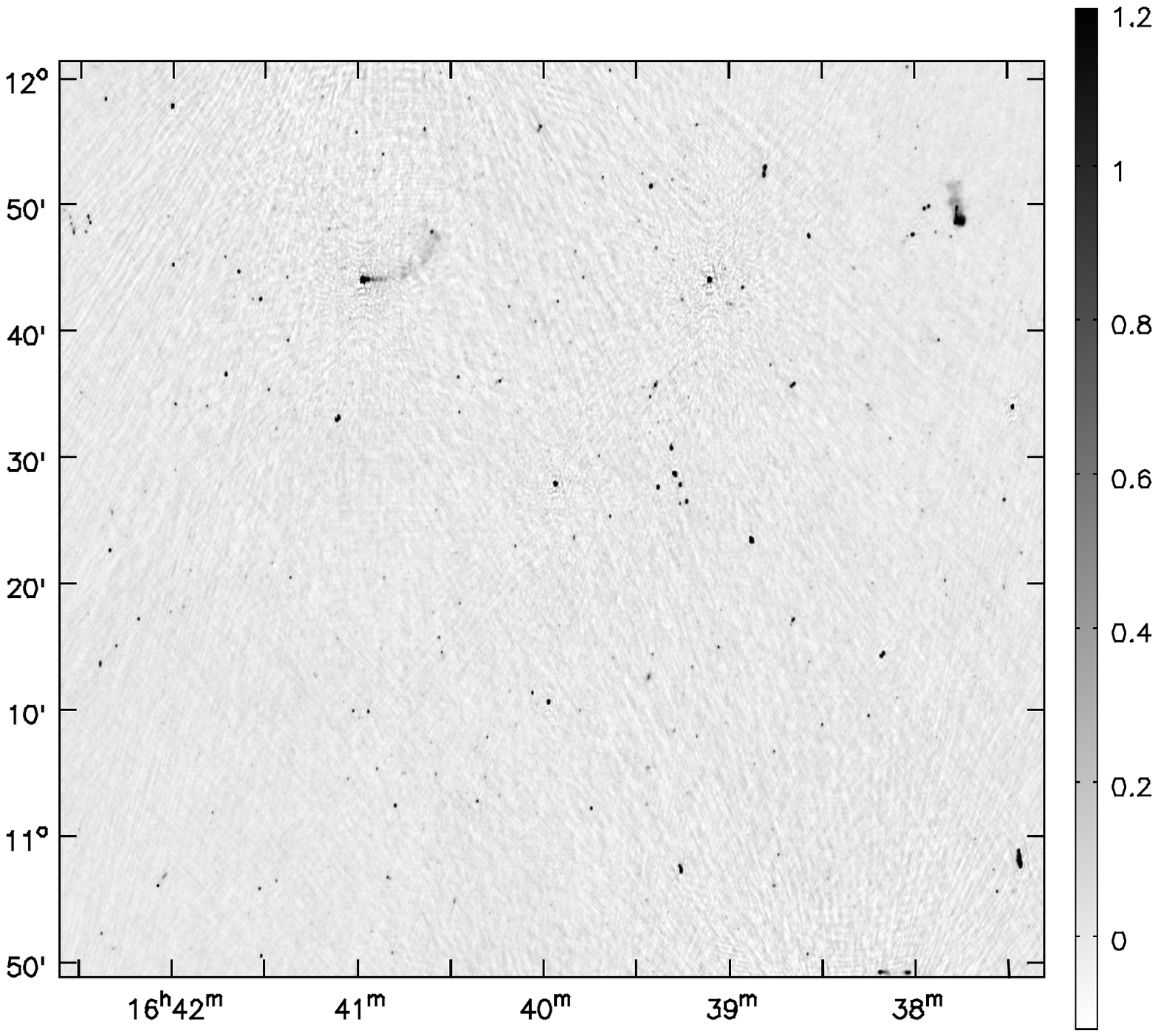}  
\includegraphics[trim = {0.0cm 5.5cm 0.0cm 3.0cm}, width=0.45\textwidth,angle=0]{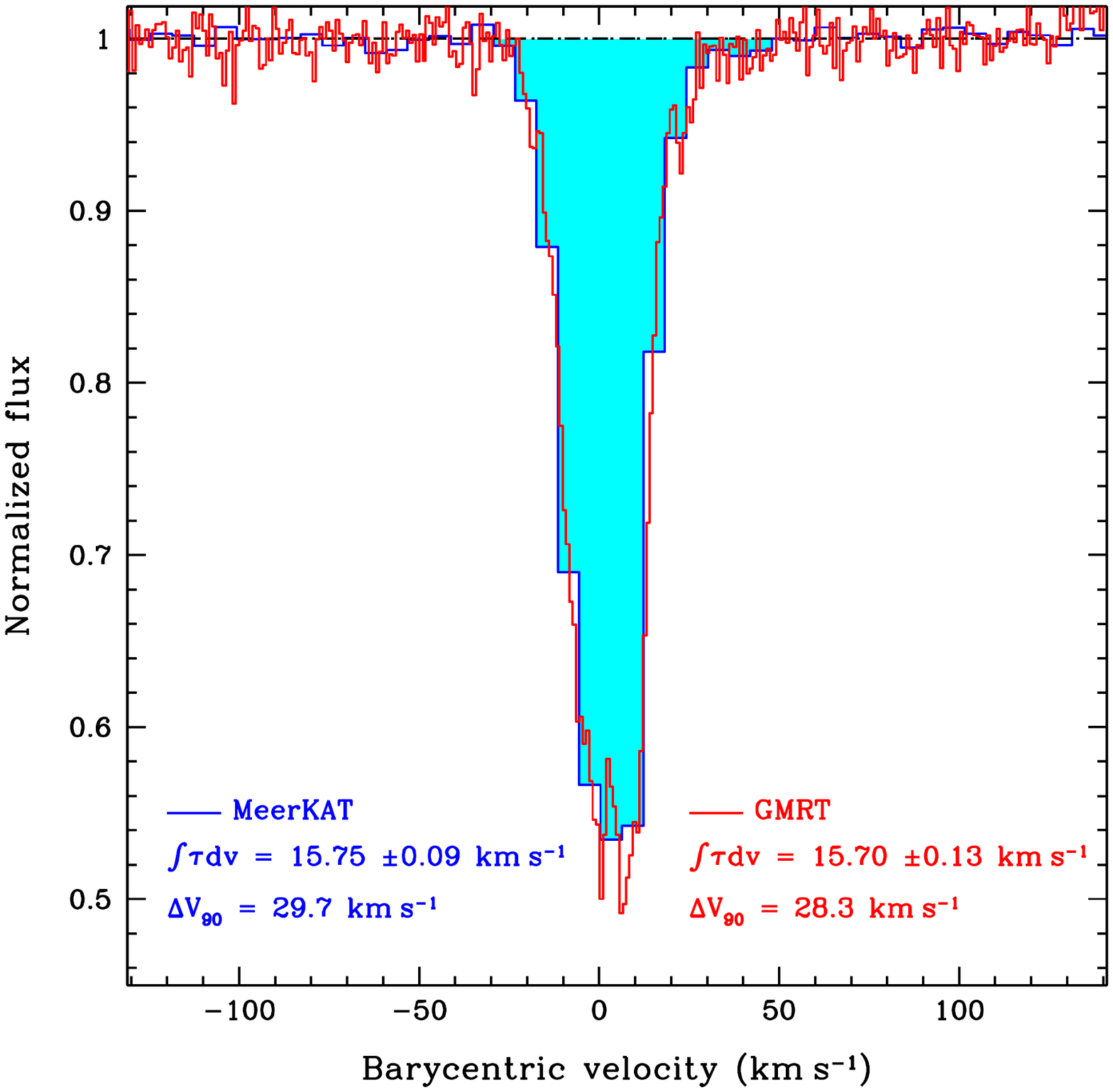}  
}} 
}}  
\vskip+0.0cm  
\caption{
	{\it Left:} A portion of the larger 3.3$^\circ$ (6k$\times$6k) MeerKAT L-band image centered on the quasar J163956.35+112758.7, with {\tt robust=0} weighting and no primary beam 
	correction applied. 
	The rms in the image is {\bf 15\,$\mu$Jy\,beam$^{-1}$}, and the restoring beam is 10.8$^{\prime\prime}\times$6.9$^{\prime\prime}$ with a position angle of -0.1$^\circ$.
	The dynamic range is $\sim$11000.
	{\it Right:} 	MeerKAT spectrum showing \hi\ 21-cm line towards J163956.35+112758.7.  The spectral rms ({\tt robust = 0}; resolution = 6\,\kms) 
	is 0.64\,mJy\,beam$^{-1}$\,channel$^{-1}$.  
	The zero of the velocity scale corresponds to $z=0.079098$.  The GMRT spectrum from 
	\citet[][]{Srianand13dib} with resolution = 0.9\,\kms\ and rms = 1.8\,mJy\,beam$^{-1}$\,channel$^{-1}$ is also shown.
} 
\label{fig:j1639}   
\end{figure} 



\end{document}